\documentclass[aps, superscriptaddress, longbibliography]{revtex4-1}

\pdfoutput=1
\usepackage{graphicx}
\usepackage{amsmath}
\usepackage{amssymb}
\usepackage{bm}
\usepackage{color}
\usepackage{comment}

\newcommand{\pa}{\partial}

\newcommand{\mean}[1]{\langle{#1}\rangle}

\newcommand{\Tr}{{\rm Tr}\hspace{0.07cm}}

\newcommand{\abs}[1]{{|#1|}}

\begin{document}
	\title{Turing instability in quantum activator-inhibitor systems}
	\author{Yuzuru Kato}
	\affiliation{Department of Complex and Intelligent Systems,
		Future University Hakodate, Hokkaido 041-8655, Japan \\
	(Corresponding author: katoyuzu@fun.ac.jp)}
	
	\author{Hiroya Nakao}
	\affiliation{Department of Systems and Control Engineering,
		Tokyo Institute of Technology, Tokyo 152-8552, Japan}
	\date{\today}

\begin{abstract}
Turing instability is a fundamental mechanism of nonequilibrium self-organization. However, despite the universality of its essential mechanism, Turing instability has thus far been investigated mostly in classical systems. In this study, we show that Turing instability can occur in a quantum dissipative system and analyze its quantum features such as entanglement and the effect of measurement. We propose a degenerate parametric oscillator with nonlinear damping in quantum optics as a quantum activator-inhibitor unit and demonstrate that a system of two such units can undergo Turing instability when diffusively coupled with each other. The Turing instability induces nonuniformity and entanglement between the two units and gives rise to a pair of nonuniform states that are mixed due to quantum noise. Further performing continuous measurement on the coupled system reveals the nonuniformity caused by the Turing instability. Our results extend the universality of the Turing mechanism to the quantum realm and may provide a novel perspective on the possibility of quantum nonequilibrium self-organization and its application in quantum technologies. 
\end{abstract}

\maketitle


\section{Introduction}

Nature displays a variety of orders that are self-organized via spontaneous symmetry breaking caused by internal interactions within systems, such as spontaneous magnetization, crystal growth, and superconductivity~\cite{camazine2003self, haken2006information, heylighen2001science}.
In particular, nonequilibrium open systems can support a wide variety of self-organized patterns that cannot occur in equilibrium systems, called dissipative structures.
Examples of dissipative structures include fluid convection patterns, laser oscillations, chemical waves and patterns, and biological patterns and rhythms~\cite{kuramoto1984chemical, nicolis1977self, prigogine1971biological}. 
Self-organization and pattern formation have also been studied in quantum systems such as atomic Bose-Einstein condensates and trapped ions~\cite{zhang2020pattern,lee2011pattern},
optomechanical systems~\cite{ludwig2013quantum}, and quantum dots~\cite{tersoff1996self}.
Quantum synchronization
\cite{lee2013quantum, lee2014entanglement, walter2014quantum, walter2015quantum, lorch2016genuine,  xu2014synchronization, roulet2018synchronizing, kato2019semiclassical, laskar2020observation, koppenhofer2020quantum, cabot2019quantum, galve2017quantum}, which has recently gained growing interest, is also an example of quantum non-equilibrium self-organization.

In 1952, Turing showed that the difference between the diffusivities of reacting chemical species can destabilize uniform stationary states and cause spontaneous emergence of nonuniform periodic patterns in spatially extended systems~\cite{turing1952chemical}.
In 1972, Gierer and Meinhardt provided an intuitive explanation of Turing instability by introducing the now well-known concept of \textit{activator-inhibitor systems} with \textit{local self-enhancement and long-range inhibition}~\cite{gierer1972theory}.
Later, Turing instability and the resulting patterns were studied in various systems, such as those undergoing chemical reactions~\cite{prigogine1968symmetry, epstein1996nonlinear, tompkins2014testing} or biological morphogenesis~\cite{meinhardt2000pattern, maini2006turing,newman2007activator}, ecological populations~\cite{mimura1978diffusive, maron1997spatial, baurmann2007instabilities}, 
and nonlinear optical systems~\cite{lugiato1987spatial, gatti1995quantum, lugiato1992quantum, zambrini2002macroscopic, lugiato1993spatial, oppo1994formation, gatti1997langevin}.
Turing patterns have also been theoretically investigated in stochastic systems~\cite{biancalani2010stochastic, butler2011fluctuation, biancalani2017giant, karig2018stochastic} and networked systems~\cite{othmer1971instability, othmer1974non, nakao2010turing, petit2017theory, muolo2019patterns}.
The first experimental realization of Turing patterns was achieved in 1990~\cite{castets1990experimental}, 40 years after Turing's seminal paper, followed by the first experimental determination of the bifurcation diagram~\cite{ouyang1991transition}, using the chlorite-iodide-malonic acid reaction in a gel reactor.
Recent progress and modern discussions on Turing instability
	have been reviewed, e.g.,  in Ref.~\cite{krause2021introduction}, and
	include various new aspects of Turing patterns including instability in multi-species systems~\cite{klika2012influence, korvasova2015investigating}, influences of domain growth~\cite{madzvamuse2010stability, klika2017history, van2021turing,klika2018domain}, and effects of delay and noise~\cite{otto2017delay}.

Recent developments in nanotechnology have stimulated both theoretical and experimental investigations of Turing-type instability and patterns in micro- and nanoscale systems, such as  rogue waves in a cavity with quantum dot molecules ~\cite{eslami2017optical}, vectorial Kerr medium \cite{zambrini2000quantum}, intracavity second harmonic generation \cite{bache2002quantum}, longitudinal microresonators \cite{chembo2016quantum},
Kerr-active microresonators~\cite{bao2020turing}, semiconductor microcavities~\cite{ardizzone2013formation}, and 
a bismuth monolayer~\cite{fuseya2021nanoscale}.
Therefore, systematic analysis of the possibility of Turing instability in quantum systems is becoming important.
In this research direction, pioneering studies on nonlinear optical systems, e.g., optical parametric oscillators \cite{lugiato1993spatial, oppo1994formation, gatti1997langevin}, have considered the possibility of pattern formation via Turing-type instability~\cite{lugiato1987spatial} and discussed the effects of quantum fluctuations~\cite{gatti1995quantum} and quantum squeezing~\cite{lugiato1992quantum}.
However, due to the difficulty in handling an infinite hierarchy of equations for operator products, the analysis was limited to the case that can be treated via the approximate stochastic differential equation of classical fields subjected to quantum fluctuations~\cite{zambrini2002macroscopic}. 

Recently, using a fully quantum-mechanical master equation, the bifurcation in a system of a pair of coupled quantum Stuart-Landau oscillators from the uniform amplitude-death state to the nonuniform oscillation-death state was discussed~\cite{bandyopadhyay2020quantum, bandyopadhyay2021quantum, bandyopadhyay2021revival}, which can be regarded as a quantum manifestation of the Turing-type bifurcation originally analyzed in a classical system~\cite{koseska2013transition}.
Though this bifurcation is interesting, it is not exactly the Turing instability in the original sense because the considered system is not of the activator-inhibitor type and does not possess a homogeneous stationary state when the coupling is absent, as discussed in Ref.~\cite{koseska2013transition}.
Additionally, the relation between the Turing bifurcation and quantum features, such as quantum entanglement and quantum measurement, has not been studied 
in these papers ~\cite{bandyopadhyay2020quantum, bandyopadhyay2021quantum, bandyopadhyay2021revival}. 

\begin{figure} [htbp]
	\begin{center}
		\includegraphics[width=0.95\hsize,clip]{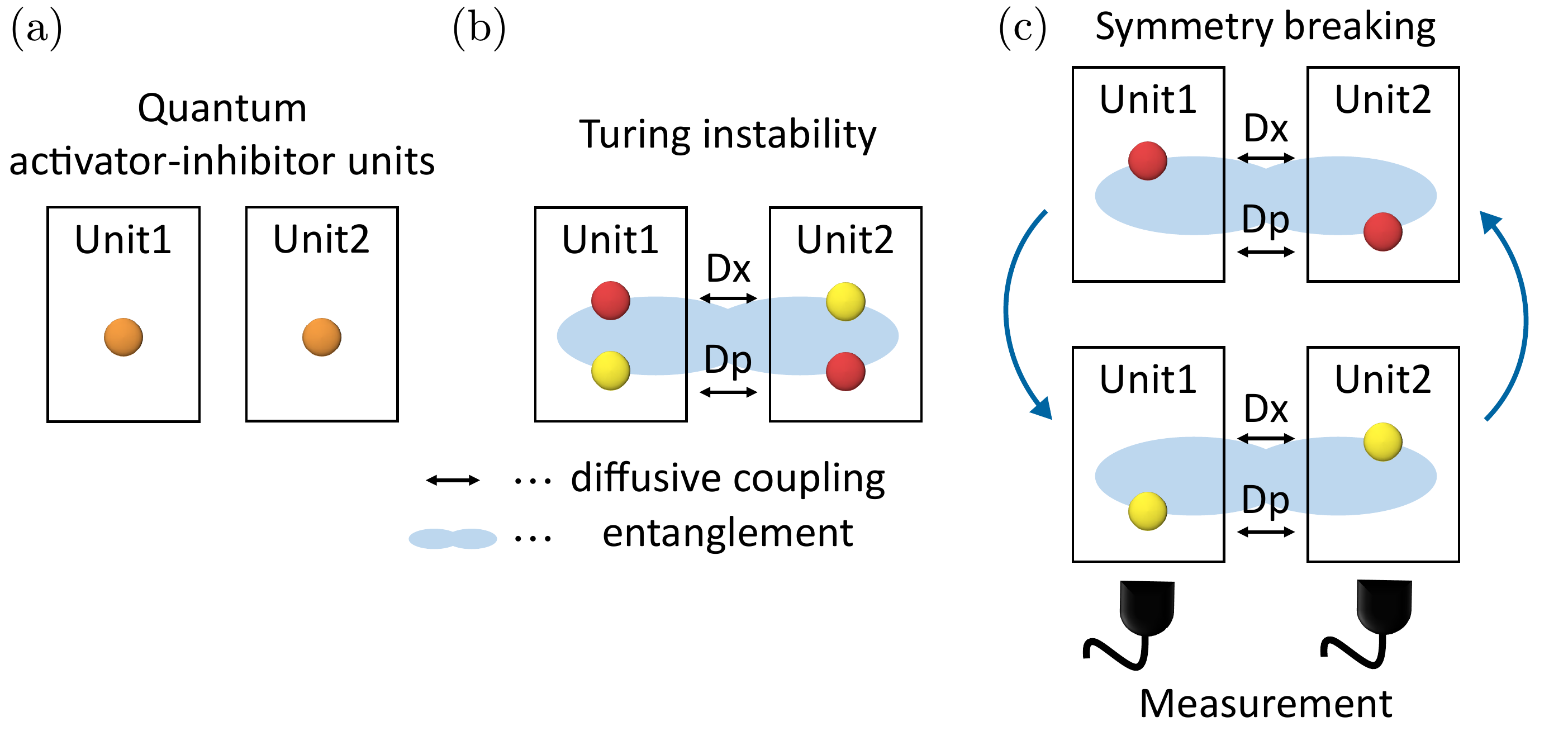}
		\caption{
			\textbf{Quantum Turing instability.} (a) Pair of quantum activator-inhibitor units. (b) Diffusive coupling between the two units can induce Turing instability, which leads to nonuniformity and entanglement between the units and yields a pair of nonuniform states that are symmetrically mixed due to quantum noise. (c) Further performing continuous measurement on the two units can break the symmetry and reveal the asymmetry caused by the Turing instability.}
		\label{fig_1}
	\end{center}
\end{figure}

In this study, we analyze Turing instability in the original sense of Turing~\cite{turing1952chemical} and Gierer and Meinhardt~\cite{gierer1972theory} in quantum dissipative systems in the simplest setting, i.e., in a pair of symmetrically coupled units, by providing a minimal model of quantum activator-inhibitor systems. 
We show that a degenerate parametric oscillator with nonlinear damping can behave as a quantum activator-inhibitor unit and that diffusive coupling between two such units can induce Turing instability and lead to nonuniformity and entanglement between the two units, which gives rise to a pair of nonuniform states that are symmetrically mixed due to quantum noise.
We further demonstrate that performing continuous measurement on the coupled system breaks this symmetry and reveals the true asymmetry caused by the Turing instability.
A schematic diagram is shown in Fig.~\ref{fig_1}.

\section{Quantum activator-inhibitor system}

\subsection{Quantum activator-inhibitor unit}

We first show that a single-mode, degenerate parametric oscillator with nonlinear damping in quantum optics~\cite{tezak2017low} can be considered a {\it quantum activator-inhibitor unit } in the sense that the deterministic trajectory of the system in the classical limit obeys conventional activator-inhibitor dynamics.

We denote by $\omega_{0}$ the resonance frequency of the cavity and by $\omega_{p}$ the frequency of the pump beam of squeezing. 
In the rotating coordinate frame of frequency $\omega_{p}/2$, 
the evolution of the density operator $\rho$ representing the system state obeys the quantum master equation (QME)~\cite{tezak2017low}
\begin{align}
	\label{eq:1s_me}
	\dot{\rho} = 
	-i \left[  \Delta a^{\dag}a 
	+ i \eta ( a^2 e^{-i \theta} - a^{\dag 2} e^{ i \theta}  )
	, \rho \right]
	+ \gamma_{1} \mathcal{D}[a]\rho + \gamma_{2}\mathcal{D}[a^{2}]\rho,
\end{align}
where 
$[A, B] = AB - BA$
is the commutator of two operators $A$ and $B$,
$a$ is the annihilation operator that subtracts a photon from the system,
$a^{\dag}$ is the creation operator that adds a photon to the system
($\dag$ denotes the Hermitian conjugate),
$\Delta = \omega_{0} - \omega_{p}/2$ is the detuning of
the resonance frequency of the system from the half frequency of the pump beam,
$\eta e^{ i \theta}$ ($\eta \geq 0$) is the squeezing parameter representing the effective amplitude of the pump beam, 
$\mathcal{D}[L]\rho = L \rho L^{\dag} - (\rho L^{\dag} L - L^{\dag} L \rho)/2$ is the Lindblad form representing the coupling of the system with the reservoirs through the operator $L$ ($L=a$ or $L=a^2$), 
and $\gamma_{1}~(>0)$ and $\gamma_{2}~(>0)$ are the decay rates for 
linear and nonlinear damping, i.e., the single-photon and two-photon loss, respectively, due to coupling of the system with the respective reservoirs. The reduced Planck constant is set as $\hbar = 1$.

\begin{figure} [htbp]
	\begin{center}
		\includegraphics[width=0.95\hsize,clip]{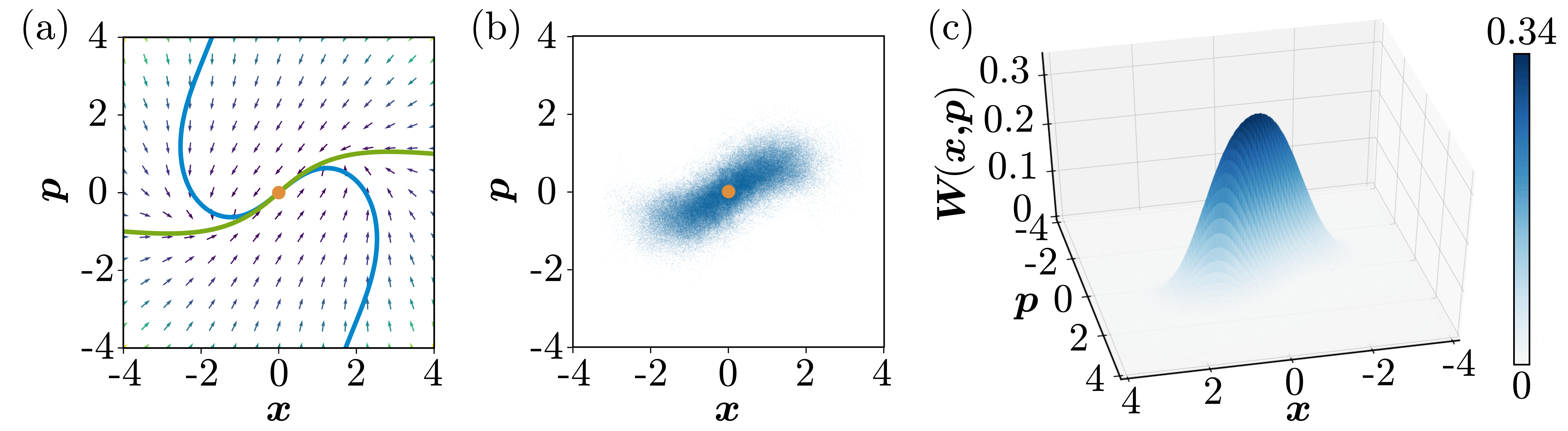}
		\caption{
			\textbf{Quantum activator-inhibitor unit.} (a) Nullclines of the deterministic vector field of Eq.~(2). Blue and green curves indicate the sets $(x, p)$ satisfying $\dot{x} = 0$ and $\dot{p} = 0$, respectively.  (b) Stochastic trajectory of $(x, p)$ obtained from the semiclassical SDE. (c) Stationary Wigner distribution $W(x, p)$ obtained from the QME.
			The parameters are $\Delta = -0.6, \gamma_{1} = 0.4, \gamma_{2} = 0.1,  \theta = \pi$, and $\eta = 0.3$.}
		\label{fig_2}
	\end{center}
\end{figure}

We employ the phase-space method~\cite{gardiner1991quantum, carmichael2007statistical} and use the Wigner distribution 
$W(x, p)$ as the quasiprobability distribution to represent the density operator $\rho$,
where $x$ and $p$ denote the position and momentum in the phase space, respectively.
	Using this approach, we can transform the QME to the evolution equation for $W(x, p)$ on the phase space, which generally has derivative terms higher than the second order. When $\gamma_2$ is small, we can neglect the higher order derivative terms, and the evolution equation for $W(x, p)$ corresponding to QME~(\ref{eq:1s_me}) can be approximated by a semiclassical Fokker-Planck equation (FPE) or the corresponding stochastic differential equation (SDE). The deterministic trajectory in the classical limit of QME~(\ref{eq:1s_me}), which neglects the effect of small quantum noise and is given by the deterministic part of the SDE,  is found to obey the following two-dimensional system:
\begin{align}
	\label{eq:1s_ctraj}
	\left( \begin{matrix}
		\dot{x}  \\
		\dot{p}  \\
	\end{matrix} \right)
	&=
	\left( \begin{matrix}
		\frac{ 2 \gamma_{2} - \gamma_{1}}{2} x + \Delta  p  
		- \gamma_{2} x  (x^{2} + p^{2}) 
		- 2 \eta ( x \cos \theta + p \sin \theta)  
		\\
		- \Delta  x + \frac{ 2 \gamma_{2} - \gamma_{1}}{2} p    
		- \gamma_{2} p  (x^{2} + p^{2}) 
		+ 2  \eta ( - x \sin \theta + p \cos \theta ) 
		\\
	\end{matrix} \right).
\end{align}
See Methods  for the detailed derivation of the equations and characterization of the quantum regime. 

By appropriately choosing the parameters, classical system~(\ref{eq:1s_ctraj}) 
obeys activator-inhibitor dynamics (see Methods).
We set the parameters such that the position $x$ and momentum $p$ play the roles of the activator and inhibitor variables, respectively, namely, $x$ autocatalytically enhances its own production while $p$ suppresses the  growth of $x$.
It is noted that the system without nonlinear damping 
can also behave as a quantum activator-inhibitor unit, but nonlinear damping is necessary to prevent the system state
from diverging to infinity after destabilization at the origin.

Figure~\ref{fig_2}(a) shows the deterministic vector field of Eq.~(\ref{eq:1s_ctraj}), where the two curves represent nullclines of $x$ and $p$ (on which $\dot{x} = 0$ or $\dot{p} = 0$) and their intersection at $(x,p) = (0, 0)$ corresponds to a stable fixed point.
Figure~\ref{fig_2}(b) shows a scatter plot of a single trajectory of the semiclassical SDE obtained by direct numerical simulations (DNSs) in the steady state (see Methods), and Fig.~\ref{fig_2}(c) shows the stationary Wigner distribution obtained from QME~(\ref{eq:1s_me}).
The semiclassical trajectory and the Wigner distribution are distributed around the classical fixed point at the origin due to quantum noise.

\subsection{Diffusively coupled quantum activator-inhibitor units}

In the classical Turing instability, the uniform stationary state of spatially distributed activator-inhibitor systems is destabilized when diffusion of the activator and inhibitor species with appropriate diffusivity is introduced, leading to the formation of nonuniform states~\cite{turing1952chemical}. 
In the simplest setting, this counterintuitive Turing instability can already be observed in a system consisting of two diffusively coupled activator-inhibitor units with identical properties: a uniform stationary state of the system, in which the two units take the same states, becomes destabilized  when the diffusivities are appropriately chosen, resulting in the formation of a nonuniform stationary state, in which the two units settle into different states from each other.

As a  quantum model that undergoes Turing instability, we diffusively couple two identical quantum activator-inhibitor units (denoted $1$ and $2$), each of which obeys Eq.~(\ref{eq:1s_me}).
The coupled system of the two units is described by a two-mode density operator $\rho$, which obeys the QME
\begin{align}
	\label{eq:2s_me}
	\dot{\rho}
	& = 
	\sum_{j=1,2} \left( -i 
	\left[\Delta a_{j}^{\dag}a_{j}
	+ i \eta ( a_{j}^2 e^{-i \theta} - a_{j}^{\dag 2} e^{ i \theta})
	, \rho \right]
	+ \gamma_{1}\mathcal{D}[a_{j}]\rho 
	+ \gamma_{2}\mathcal{D}[a_{j}^{2}]\rho 
	\right)
	\cr
	& -i \left[ i \frac{D_{h}}{4} \left\{ (a_1 - a_2)^2 - (a_1^\dag - a_2^\dag)^2 \right\}, \rho \right]
	+ 
	D_{c} \mathcal{D}[a_{1} - a_{2}]\rho,
\end{align}
where $a_j$ and $a_j^{\dag}$ are the annihilation and creation operators for the $j$th quantum activator-inhibitor unit ($j = 1, 2$), respectively.
The parameters $\Delta, \eta e^{ i \theta}, \gamma_1$ and $\gamma_2$
are common to both  units.
In this equation, the first line represents the two single-mode units given by Eq.~(\ref{eq:1s_me}), and the newly introduced terms in the second line represent the coupling between the two units.
The first coupling term can be represented as a sum of squeezing terms, i.e.,
$- i \left[ i \frac{D_{h}}{4} \left\{ (a_1 - a_2)^2 \right. \right.$ $\left. \left. -  (a_1^\dag - a_2^\dag)^2 \right\}, \rho \right] = 
\sum_{j=1,2} \left(-i  \left[ i \frac{D_{h}}{4} ( a_{j}^2  -  a_{j}^{\dag 2}), \rho \right] \right)-i \left[i \frac{D_{h}}{2} ( a_{1}^{\dag} a_{2}^{\dag} - a_{1} a_{2}), \rho \right]$,
which can be interpreted as single-mode and two-mode squeezing Hamiltonians, respectively. The second term with $D_{c}$ represents dissipative coupling, namely, a coupling arising from dissipative processes~\cite{lee2014entanglement, walter2015quantum}.
It is noted that Eq.~(\ref{eq:2s_me}) is symmetric with respect to the exchange of the units $1$ and $2$.

By employing the phase-space method for two-mode systems, the deterministic dynamics in the classical limit of QME~(\ref{eq:2s_me}) can be derived as
(see Methods)
\begin{align}
	\label{eq:2s_ctraj}
	&\left( \begin{matrix}
		\dot{x}_1  \\
		\dot{p}_1  \\
		\dot{x}_2  \\
		\dot{p}_2  \\
	\end{matrix} \right)=
	&
	\left( \begin{matrix}
		\frac{2 \gamma_{2} - \gamma_{1}}{2} x_1  + \Delta  p_1  
		- \gamma_{2} x_1  (x_1^{2} + p_1^{2}) 
		- 2 \eta ( x_1 \cos \theta + p_1 \sin \theta)  
		+ D_x(x_2 - x_1)
		\\
		- \Delta  x_1 + \frac{2 \gamma_{2} - \gamma_{1}}{2} p_1    
		- \gamma_{2} p_1  (x_1^{2} + p_1^{2}) 
		+ 2  \eta ( - x_1 \sin \theta + p_1 \cos \theta ) 
		+ D_{p}(p_2 - p_1)
		\\
		\frac{2 \gamma_{2} - \gamma_{1}}{2} x_2  + \Delta  p_2  
		- \gamma_{2} x_2  (x_2^{2} + p_2^{2}) 
		- 2 \eta ( x_2 \cos \theta + p_2 \sin \theta)  
		+ D_x(x_1 - x_2)
		\\
		- \Delta  x_2 + \frac{2 \gamma_{2} - \gamma_{1}}{2} p_2  
		- \gamma_{2} p_2  (x_2^{2} + p_2^{2}) 
		+ 2  \eta ( - x_2 \sin \theta + p_2 \cos \theta ) 
		+ D_{p}(p_1 - p_2)
		\\
	\end{matrix} \right),
\end{align}
where $x_j$ and $p_j$ represent the position and momentum of the $j$th unit in the phase space of the two-mode Wigner distribution $W(x_1, p_1, x_2, p_2)$~\cite{carmichael2007statistical}.
We see that two classical activator-inhibitor units, each of which is described by Eq.~(\ref{eq:1s_ctraj}), are diffusively coupled through the position $x$ (activator) and momentum $p$ (inhibitor) by the last term in each equation. 
These terms arise from the single- and two-mode squeezing Hamiltonians whose intensities are characterized by $D_h$
and from the dissipative coupling whose intensity is characterized by $D_c$ in Eq.~(\ref{eq:2s_me}). The diffusion constants of $x$ and $p$ in Eq.~(\ref{eq:2s_ctraj}) are given by $D_x = (D_{c} + D_{h})/2$ and $D_p = (D_{c} - D_{h})/2$, respectively.
It should be noted that the first  term characterized by $D_h$ represents a Hamiltonian coupling and non-dissipative, but it acts as a dissipative coupling in the deterministic dynamics in the classical limit in Eq.~(\ref{eq:2s_ctraj}).

The classical coupled system described by Eq.~(\ref{eq:2s_ctraj}) can undergo Turing instability when the conditions of \textit{local self-enhancement} and \textit{long-range inhibition} are satisfied (see Methods).
Therefore, the quantum activator-inhibitor system, Eq.~(\ref{eq:2s_me}), is also expected to exhibit Turing instability when the parameter values are appropriately chosen.
Our aim in this study is to clarify whether Turing instability can occur within the original activator-inhibitor framework in the simplest setting in quantum dissipative systems.
We note that the requirements of a coupled activator-inhibitor pair or the existence of homogeneous solution can be relaxed when we consider more general models~\cite{madzvamuse2010stability, klika2017history, van2021turing,klika2012influence, korvasova2015investigating,klika2018domain,otto2017delay}.
In this study, we focus on the simplest case of a pair of symmetrically coupled quantum activator-inhibitor units and discuss quantum Turing instability in the original sense of Turing~\cite{turing1952chemical} and Gierer-Meinhardt~\cite{gierer1972theory}. Due to its simplicity, the model allows the direct numerical simulations of quantum dynamics and is the most amenable to experiment.


\section{Turing instability}

\subsection{Semiclassical regime}
Deterministic system (\ref{eq:2s_ctraj}) has a fixed point at the origin of the 4-dimensional phase space, i.e., $(x_1, p_1, x_2, p_2) = (0, 0, 0, 0)$, which is stable when diffusive coupling is absent, i.e., $D_x = D_p = 0$. Both units $1$ and $2$ settle to the origin, i.e., $(x_j, p_j) = (0, 0)$ for $j=1, 2$; hence, the whole system takes a uniform state.
When diffusive coupling with appropriate diffusivities is introduced, this uniform state
is destabilized by the Turing instability, and instead, a pair of stable nonuniform fixed points appear
at $(x_1, p_1, x_2, p_2) = (\pm A,\pm B,\mp A, \mp B)$ of deterministic classical system (\ref{eq:2s_ctraj})  (see Methods).

Correspondingly, in quantum system~(\ref{eq:2s_me}), when the diffusive coupling is absent ($D_x = D_p = 0$),  the state of each unit localizes around the stable fixed point at $(0, 0)$ as shown in Fig.~\ref{fig_2}(a). Thus, the two units obey the same distribution and the whole system
is in the uniform state. 
However, when the diffusion constants are appropriately chosen, this uniform state is destabilized by the Turing instability and gives way to nonuniform states as demonstrated below.

\begin{figure} [htbp]
	\begin{center}
		\includegraphics[width=0.95\hsize,clip]{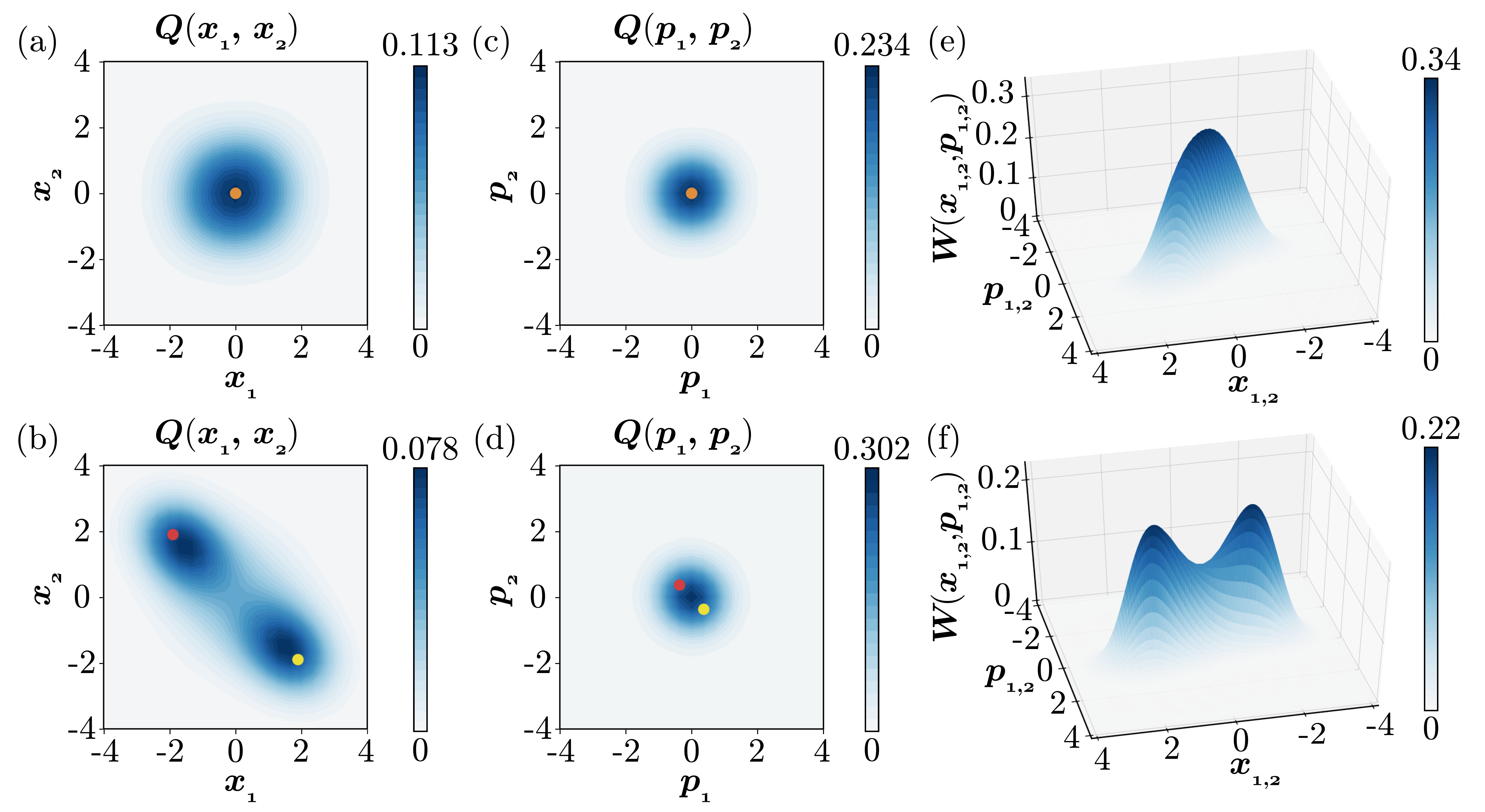}
		\caption{
			\textbf{Turing instability in a pair of diffusively coupled quantum activator-inhibitor units in the semiclassical regime.}
			(a, b) 2D plots of the Q distribution $Q(x_1, x_2)$. 	
			(c, d) 2D plots of the Q distribution $Q(p_1, p_2)$. 
			(e, f) 3D plots of the stationary Wigner distributions 
			$W(x_1, p_1)$ and $W(x_2, p_2)$ of the units $1$ and $2$.
			Red and yellow dots in (a-d) represent stable fixed points of the deterministic system in the classical limit.
			In (a, c, e), the two units are uncoupled. The states of the units are uncorrelated and localized around the origin;
			hence, the whole system is in a uniform state.
			In (b, d, f), the two units are diffusively coupled. Due to the Turing instability, the two units tend to take different states from each other; hence, the whole system is nonuniform.
			In (e, f), the Wigner distributions for the units $1$ and $2$ are identical to each other and hence shown as a single plot.
			The parameters of the quantum activator-inhibitor units are $\Delta = -0.6, \gamma_{1} = 0.4, \gamma_{2} = 0.1, \theta = \pi$, and $\eta = 0.3$.
			The diffusion constants are $D_x = D_p = 0$ ($D_h = 0$ and $D_c = 0$) in (a, c, e) and $D_x = 0.005$ and $D_p = 0.995$ ($D_h = -0.99$ and $D_c = 1$) in (b, d, f).
		}
		\label{fig_3}
	\end{center}
\end{figure}

Figure~\ref{fig_3} shows the Turing instability in the semiclassical regime observed by DNSs of QME~(\ref{eq:2s_me}). 
The same parameters as in Fig.~\ref{fig_2} are assumed for both units.
The two units are uncoupled ($D_x = D_p = 0$) in Figs.~\ref{fig_3}(a,~c,~e), while they are coupled with appropriate diffusion constants ($D_x = 0.005, D_p = 0.995$) in Figs.~\ref{fig_3}(b,~d,~f).
To visualize the nonuniformity of the system state $\rho$, we introduce the two-mode Husimi Q distribution \cite{gardiner1991quantum, carmichael2007statistical} 
$Q\left(x_{1}, p_{1}, x_{2}, p_{2}\right) =\frac{1}{\pi^{2}}  \left\langle \alpha_1, \alpha_2 | \rho | \alpha_1, \alpha_2 \right\rangle$ with $\alpha_j = x_j + i p_j~(j = 1, 2)$ and use the marginal distributions 
$Q(x_1,x_2) =  \int \int dp_1 dp_2 Q\left(x_{1}, p_{1}, x_{2}, p_{2}\right)$ and $Q(p_1,p_2) =  \int \int dx_1 dx_2 Q\left(x_{1}, p_{1}, x_{2}, p_{2}\right)$ of the position (activator) variables $x_{1,2}$ and momentum (inhibitor) variables $p_{1,2}$ calculated from $Q\left(x_{1}, p_{1}, x_{2}, p_{2}\right)$.

In Figs.~\ref{fig_3}(a, c) without diffusive coupling, both $Q(x_1, x_2)$ and $Q(p_1, p_2)$ are symmetrically distributed around the origin.
The variables of the two units are uncorrelated and statistically exhibit the same distribution.
Thus, the state $\rho$ of the whole system consisting of the two units is symmetric and uniform.
In contrast, in Figs.~\ref{fig_3}(b, d) with diffusive coupling, $Q(x_1, x_2)$ is not symmetric and takes two extrema near the two classical fixed points $(x_1, x_2) = (A, -A)$ and $(-A, A)$, and similarly $Q(p_1, p_2)$ takes two extrema near $(p_1, p_2) = (B, -B)$ and $(-B, B)$.
Thus, the two units tend to take the opposite states from each other and the state $\rho$ of the whole system is nonuniform.
It is noted that, because of quantum noise, the system state is mixed and the distributions have two symmetric peaks near both of the classical fixed points.

Figures~\ref{fig_3}(e,~f) show the marginal Wigner distributions $W(x_1, p_1)$ and $W(x_2, p_2)$ of units 1 and 2 for the cases without (e) and with (f) diffusive coupling.
These Wigner functions are obtained from the marginal density operators $\rho_1 = \Tr_2[\rho]$ and $\rho_2 = \Tr_1[\rho]$, where $\Tr_j[\cdot]$ represents the partial trace over system $j$ in the semiclassical regime. 
Due to the symmetry of the two units, $W(x_1, p_1)$ and $W(x_2, p_2)$ are identical to each other.
Additionally, the Wigner distributions in Fig.~\ref{fig_3}(e) without diffusive coupling are identical to that of a single unit shown in Fig.~\ref{fig_2}(c).
In Fig.~\ref{fig_3}(e) without diffusive coupling, the Wigner distributions have a single peak at the origin,
whereas in Fig.~\ref{fig_3}(f) with diffusive coupling, the Wigner distributions have two symmetric
peaks near the two stable fixed points $(x_1, p_1, x_2, p_2) = (\pm A,\pm B,\mp A, \mp B)$ of deterministic classical system~(\ref{eq:2s_ctraj}) (see Methods). 

The above results clearly indicate that Turing instability has indeed occurred and resulted in the formation of nonuniform stationary states in two diffusively coupled quantum
activator-inhibitor units described by Eq.~(\ref{eq:2s_me}).
In this regime, we can also perform direct numerical simulations of the corresponding SDE, which clearly visualize the 
nonuniformity caused by the Turing instability (see Methods).

\begin{figure} [htbp]
	\begin{center}
		\includegraphics[width=0.95\hsize,clip]{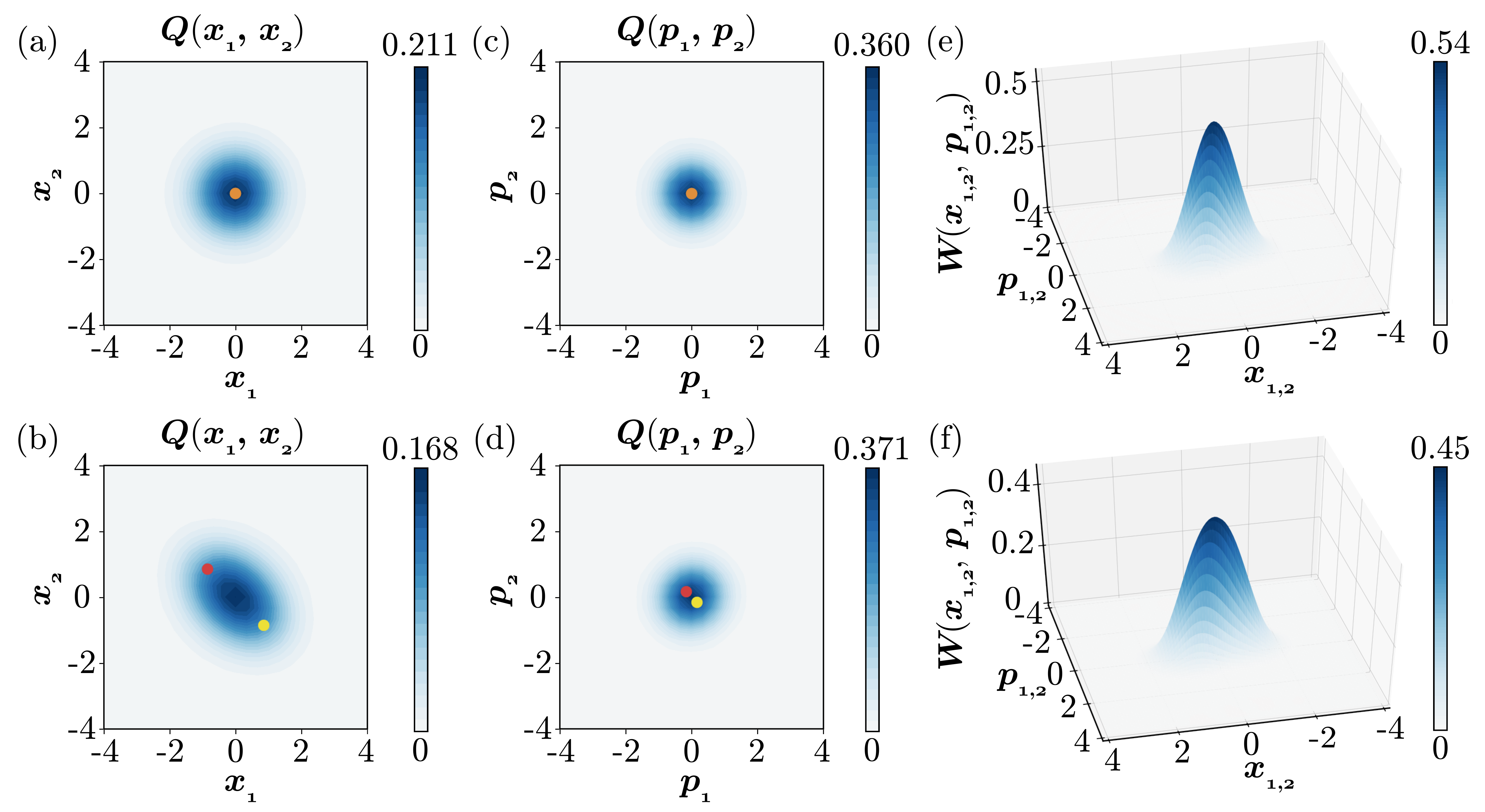}
		\caption{
			\textbf{Turing instability in a pair of diffusively coupled quantum activator-inhibitor units in the weak quantum regime.}
			(a, b) 2D plots of the Q distribution $Q(x_1, x_2)$. 	
			(c, d) 2D plots of the Q distribution $Q(p_1, p_2)$. 
			(e, f) 3D plots of the stationary Wigner distributions 
			$W(x_1, p_1)$ and $W(x_2, p_2)$ of units $1$ and $2$ (identical to each other).
			Red and yellow dots in (a-d) represent stable fixed points of the deterministic system in the classical limit.
			In (a, c, e), the two units are uncoupled. The states of the units are localized around the origin and uncorrelated with each other. In (b, d, f), the two units are diffusively coupled. Due to the Turing instability, the two units tend to take different states from each other and show a nonuniform distribution.
			The parameters of the quantum activator-inhibitor units are $\Delta = -0.6, \gamma_{1} = 1.2, \gamma_{2} = 0.5, \theta = \pi$, and $\eta = 0.3$.
			The diffusion constants are $D_x = D_p = 0$ ($D_h = 0$ and $D_c = 0$) in (a, c, e) and $D_x = 0.005$ and $D_p = 0.995$ ($D_h = -0.99$ and $D_c = 1$) in (b, d, f).
		}
		\label{fig_4}
	\end{center}
\end{figure}

\begin{figure} [htbp]
	\begin{center}
		\includegraphics[width=0.95\hsize,clip]{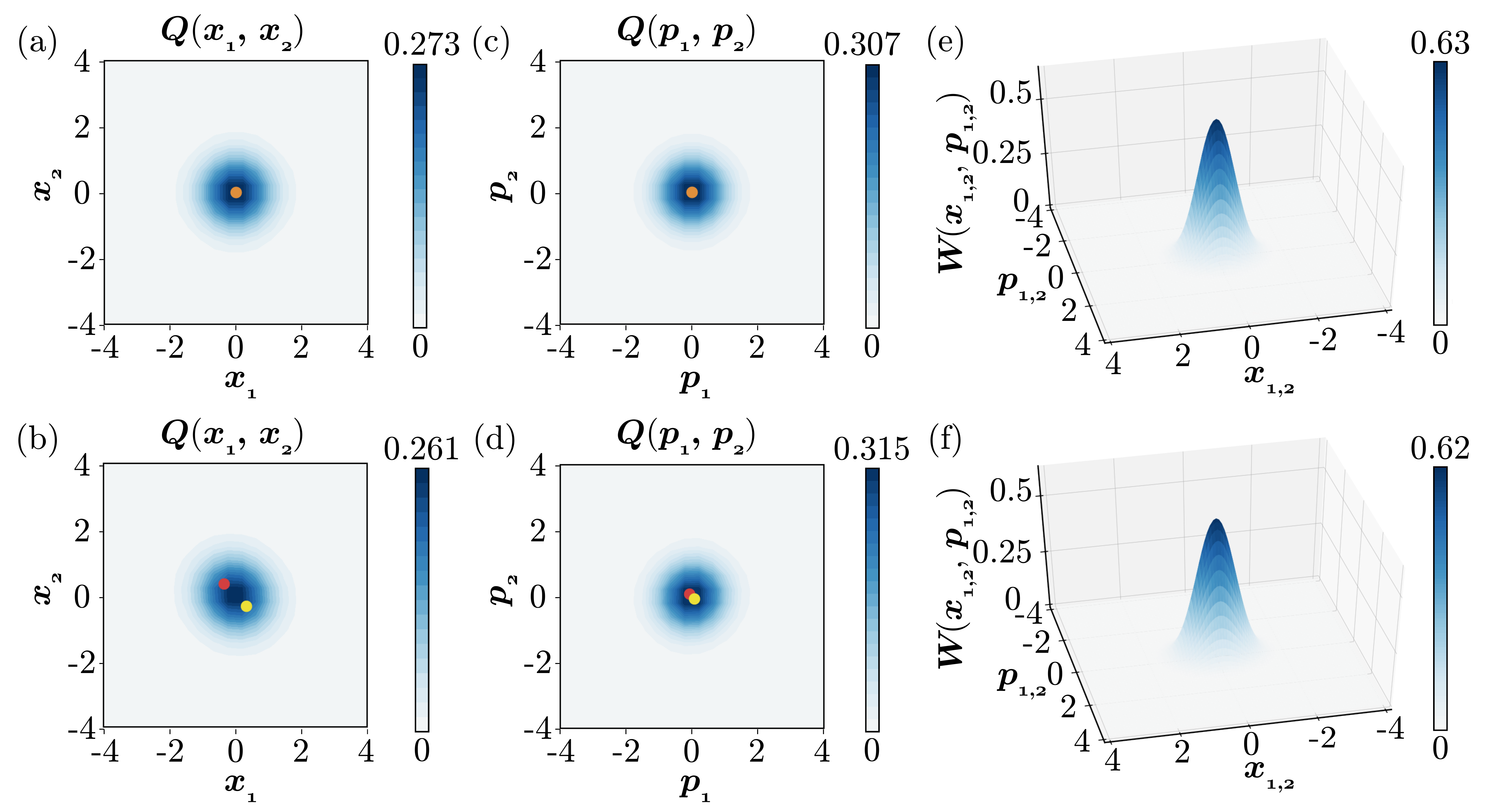}
		\caption{
			\textbf{Turing instability in a pair of diffusively coupled quantum activator-inhibitor units in the strong quantum regime.}
			(a, b) 2D plots of the Q distribution $Q(x_1, x_2)$. 	
			(c, d) 2D plots of the Q distribution $Q(p_1, p_2)$. 
			(e, f) 3D plots of the stationary Wigner distributions 
			$W(x_1, p_1)$ and $W(x_2, p_2)$ of units $1$ and $2$ (identical to each other).
			Red and yellow dots in (a-d) represent stable fixed points of the deterministic system in the classical limit.
			In (a, c, e), the two units are uncoupled. The states of the units are localized around the origin and uncorrelated with each other. In (b, d, f), the two units are diffusively coupled. Due to the Turing instability, the two units tend to take different states from each other and show a nonuniform distribution.
			The parameters of the quantum activator-inhibitor units are $\Delta = -0.6, \gamma_{1} = 6.2, \gamma_{2} = 3, \theta = \pi$, and $\eta = 0.3$.
			The diffusion constants are $D_x = D_p = 0$ ($D_h = 0$ and $D_c = 0$) in (a, c, e) and $D_x = 0.005$ and $D_p = 0.995$ ($D_h = -0.99$ and $D_c = 1$) in (b, d, f).
		}
		\label{fig_5}
	\end{center}
\end{figure}

\subsection{Weak quantum regime}

Next, we show the results for the weak quantum regime. We set the parameters of QME~(\ref{eq:2s_me}) in a deeper quantum regime while keeping the deterministic system in the classical limit, Eq.~(\ref{eq:2s_ctraj}), remain unchanged from the previous semiclassical case. See Methods for the characterization of the quantum regime.
Figure~\ref{fig_4} shows the Turing instability in this regime. The two units are uncoupled in Figs.~\ref{fig_4}(a,~c,~e), while they are coupled with appropriate diffusion constants in Figs.~\ref{fig_4}(b,~d,~f).

As in the previous semiclassical case, when diffusive coupling is absent,  
the marginal Q distributions $Q(x_1,x_2)$ and $Q(p_1, p_2)$
of activator $x$ and inhibitor $p$ are symmetrically localized around the origin in Figs.~\ref{fig_4}(a,~c).
When diffusive coupling is introduced, these joint distributions become nonsymmetric, indicating that the two units are anticorrelated and tend to take the opposite states from each other as shown in Figs.~\ref{fig_4}(b,~d).
In this regime, due to the strong nonlinear damping, the two stable fixed points in the classical limit are closer to each other than in the semiclassical regime.
Correspondingly, the nonuniformity of the joint distributions is less pronounced than in the semiclassical case due to the relatively stronger effect of quantum noise.

Figures~\ref{fig_4}(e, f) show the marginal Wigner distributions $W(x_1, p_1)$ and $W(x_2, p_2)$ of units 1 and 2, which are identical to each other, before (e) and after (f) the Turing instability.
Compared with the Wigner distribution in Fig.~\ref{fig_4}(e) before the Turing instability, the Wigner distribution in Fig.~\ref{fig_4}(f) after the instability is more elongated along the axis on which the two classical stable fixed points exist, although double symmetric
peaks as in the semiclassical case are not observed due to the strong effect of quantum noise.

Thus, although blurred by quantum noise, the system undergoes a transition from the uniform state to the nonuniform state with the introduction of diffusive coupling, namely, the Turing instability also occurs in the quantum regime considered here.

\subsection{Strong quantum regime}

We also consider a strong quantum regime with a larger decay rate for nonlinear damping. Figure~\ref{fig_5} shows the Turing instability in this regime. As the fluctuations are stronger than the two previous cases due to the effect of stronger quantum noise,  only a slight nonuniformity can be observed.
As shown later, the nonuniformity between the two units in this regime can be more clearly observed  by using continuous measurement.

\subsection{Phase diagram: nonuniformity and entanglement}

We have seen that Turing instability occurs in a pair of diffusively coupled quantum activator-inhibitor units in the semiclassical, weak quantum, and strong quantum regimes.
Here, we analyze the dependence of the system's behavior on the diffusion constants and the relationship between the Turing instability and quantum entanglement.
We use the same parameter sets for the quantum activator-inhibitor units as in Figs.~\ref{fig_3}, ~\ref{fig_4}, and ~\ref{fig_5} for the semiclassical, weak quantum, and strong quantum regimes, respectively.

\begin{figure} [htbp]
	\begin{center}
		\includegraphics[width=0.95\hsize,clip]{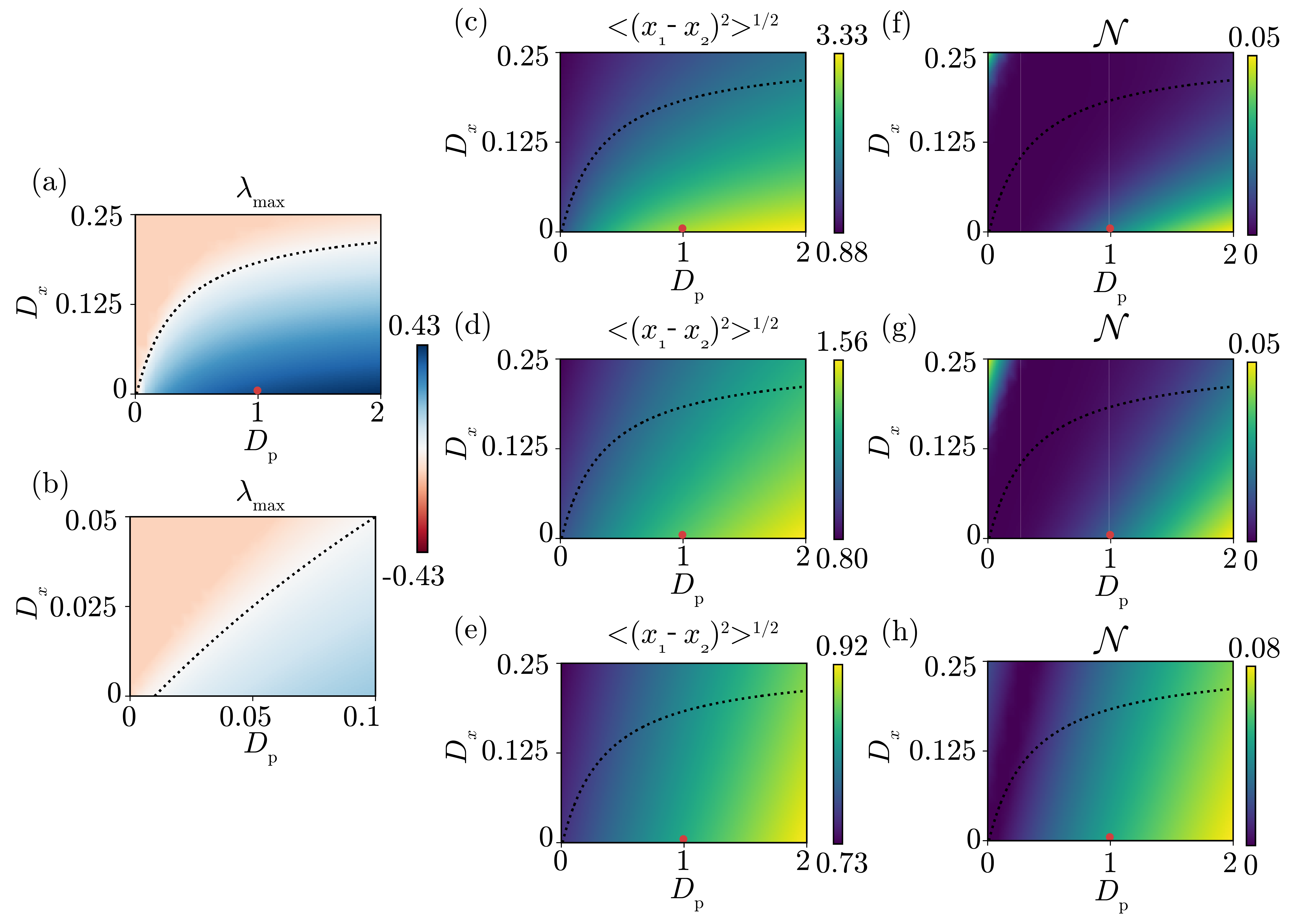}
		\caption{
			\textbf{Dependence of the eigenvalue, nonuniformity, and negativity on the diffusion constants $D_x$ and $D_p$. }
			(a, b) Maximum eigenvalues $\lambda_{max}$. (b) shows a blowup of (a) near the origin.
			(c, d, e) Root mean squared distance $\sqrt{\mean{(x_1 - x_2)^2}}$.
			(f, g, h) Negativity $\mathcal{N}$.
			In each figure, the critical curve of the Turing instability in the classical limit (i.e., on which $\lambda_{max} = 0$) is represented by a black-dotted curve and the red dot represents
			the diffusivities $(D_x, D_p) = (0.005, 0.995)$ used in Figs.~\ref{fig_3},  ~\ref{fig_4}, and ~\ref{fig_5}.			%
			The parameters are $\Delta = -0.6,  \theta = \pi$, $\eta = 0.3$, and $\frac{2 \gamma_{2} - \gamma_{1}}{2} = -0.1$,
			where $\gamma_{1} = 0.4, \gamma_{2} = 0.1$ in the semiclassical regime (c, f), 
			$\gamma_{1} = 1.2, \gamma_{2} = 0.5$ in the weak quantum regime (d, g),
			$\gamma_{1} = 6.2, \gamma_{2} = 3$ in the strong quantum regime (e, h).
		}
		\label{fig_6}
	\end{center}
\end{figure}

Figure~\ref{fig_6} plots the (i) maximum eigenvalue $\lambda_{max}$ of the linearized equation of Eq.~(\ref{eq:2s_ctraj}) in the classical limit (a,~b), (ii) root mean squared difference (RMSD) $\sqrt{\mean{(x_1 - x_2)^2}} = \sqrt{\Tr[(x_1 - x_2)^2 \rho]}$ quantifying the nonuniformity between the two units (c, d, e), and (iii) negativity ${\cal N}$ (see Methods) characterizing the degree of quantum entanglement (f,~g, h) on the $D_x - D_p$ plane.
We note that Figs.(a) and~(b) are common to all regimes, Figs. (c) and~(f) are for the semiclassical regime, Figs.~(d) and~(g) are for the weak quantum regime, and Figs.~(e) and~(h) are for the strong quantum regime. 
 
As shown in Figs.~\ref{fig_6}(a,~b), the eigenvalue $\lambda_{max}$ of the uniform state is positive in the region below the dotted curve, where the diffusivity of the inhibitor $D_p$ is relatively large compared to that of the activator $D_x$. Turing instability is expected to occur also in this region in the quantum system.
The red dot ($D_x = 0.005, D_p = 0.995$) represents the diffusion constants in the classical limit corresponding to Figs.~\ref{fig_3},  Figs.~\ref{fig_4}, and~\ref{fig_5}.

The RMSD plotted in Figs.~\ref{fig_6}(c,~d, e) shows that the nonuniformity is indeed caused by the Turing instability in the semiclassical, weak quantum, and strong quantum regimes and 
significantly correlated with the maximal eigenvalue $\lambda_{max}$ in the classical limit.
There is a tendency that
the nonuniformity is most strongly pronounced in the semiclassical regime (c), moderately in the weak quantum regime (d), and only weakly in the strong quantum regime (e) , reflecting that the quantum noise is weaker and that the system state more clearly localizes around the two classical fixed points in  this order (see Figs.~\ref{fig_3}, ~\ref{fig_4}, and  ~\ref{fig_5}).

The negativity ${\cal N}$ shown in Figs.~\ref{fig_6}(f,~g,~h) also increases with $\lambda_{max}$, indicating that quantum entanglement between the two units also arises in the nonuniform state yielded by the Turing instability.
Thus, the entanglement tends to be positively correlated with the nonuniformity between the two activator-inhibitor units and becomes stronger in the lower-right part where $D_x$ is small while $D_p$ is large in this parameter region.
It is noted that a high-${\cal N}$ region also arises when $D_p$ is close to zero while $D_x$ is relatively large, which is outside the Turing-unstable region and simply shows that the two units are already entangled before the onset of Turing instability by the effects of two-mode squeezing and dissipative coupling.

\subsection{Symmetry breaking via continuous measurement}

We have observed that Turing instability destabilizes the uniform state of the system of two units and gives rise to nonuniformity.
The distributions in the nonuniform state are localized around the two classical fixed points as observed in 
Figs.~\ref{fig_3},  ~\ref{fig_4}, and ~\ref{fig_5}.
This can be interpreted as a quantum-mechanically mixed state of the two classical situations where the system converges to either of the two stable fixed points.
Thus, in contrast to the classical Turing instability in which only one of the two states is realized depending on the initial conditions, the symmetry of the coupled system is still preserved due to quantum noise even if the system state is nonuniform.
Here, we show that further performing continuous measurement on the system can break this symmetry and reveal the true asymmetry of the system, which can be observed only in quantum systems.
A similar measurement-induced spontaneous $\mathbb{Z}_2$ symmetry breaking in a spin-chain system has been reported in Ref.~\cite{garcia2019spontaneous}.

We introduce continuous measurement on the linear damping 
(single-photon loss) bath coupled to each unit in QME~(\ref{eq:2s_me}).
The stochastic master equations (SMEs) describing the system and the measurement results are then given by \cite{wiseman2009quantum} 
\begin{align}
	\label{eq:2s_sme}
	& d{\rho}
	= 
	\left\{
	\sum_{j=1,2} \left( -i 
	\left[\Delta a_{j}^{\dag}a_{j}
	+ i \eta ( a_{j}^2 e^{-i \theta} - a_{j}^{\dag 2} e^{ i \theta})
	\rho \right]
	+ \gamma_{1}\mathcal{D}[a_{j}]\rho 
	+ \gamma_{2}\mathcal{D}[a_{j}^{2}]\rho 
	\right)
	\right.
	\cr
	&
	\left.
	- i \left[  i \frac{D_{h}}{4}\{ (a_1 - a_2)^2 - (a_1^\dag - a_2^\dag)^2 \}, \rho \right]
	+ 
	D_{c} \mathcal{D}[a_{1} - a_{2}]\rho
	\right\} dt
	+ 
	\sum_{j=1,2}\sqrt{\kappa_j \gamma_{1}}\mathcal{H}[a_j e^{- i \phi_j}]\rho dW_j,
	\cr
	&dY_j
	= \sqrt{\kappa_j \gamma_{1}}\hspace{0.05cm}{\rm Tr}[(a_j e^{- i \phi_j} 
	+ a_j^{\dag} e^{ i \phi_j})\rho]dt 
	+ dW_j,
	\quad
	(j=1, 2)
\end{align}
where the first equation describes the stochastic evolution of the density operator $\rho$ of the whole system under the effect of the measurement and the second equation describes the result $Y_j$ ($j=1, 2$) of the measurement on each unit.
The term $\mathcal{H}[L]\rho = L \rho + \rho L^{\dag} - \Tr[(L + L^\dagger) \rho]\rho$
represents the effect of measurement performed on the quadrature
$L + L^{\dag}$;
$\kappa_j$ and $\phi_j~(0 \leq \kappa_j \leq 1, 0 \leq \phi_j < 2 \pi)$ represent the efficiency and quadrature angle of the measurement on the $j$th unit $~(j = 1,2)$, respectively; $Y_j$ is the output of the measurement result on the $j$th unit $~(j = 1,2)$; and $dW_{1}$ and $dW_{2}$ represent independent Wiener processes satisfying $\mean{dW_k(t) dW_l(t)} = \delta_{kl} dt$ for $k,l = 1, 2$.
In contrast to QME, which gives averaged results over all possible measurement outcomes, this SME gives a single quantum trajectory of the system under the continuous measurement and can reveal the symmetry breaking of the system, which is preserved due to quantum noise in the steady state of QME.

\begin{figure} [htbp]
	\begin{center}
		\includegraphics[width=0.95\hsize,clip]{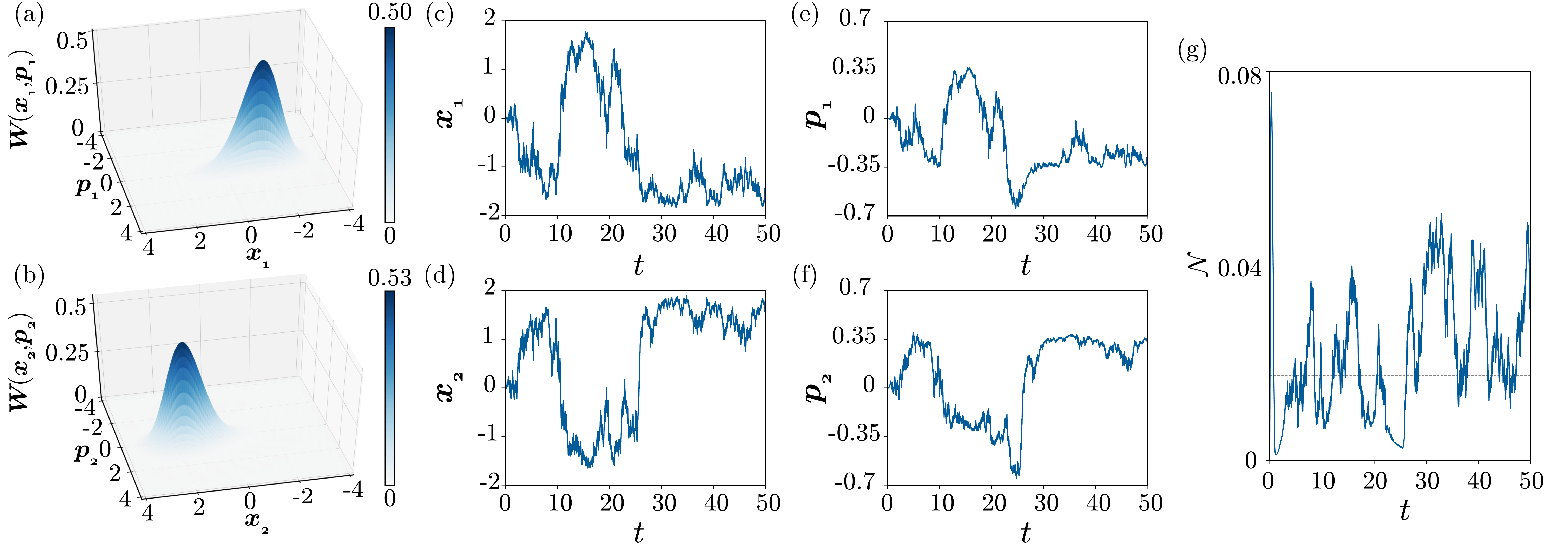}
		\caption{
			\textbf{Turing instability under continuous quantum measurement in the semiclassical regime.}
			(a, b) 3D snapshot plots of the Wigner distributions $W(x_1, p_1)$ and $W(x_2, p_2)$ at $t = 50$.
			(c, d, e, f) Time evolution of the average values of the position and momentum
			operators for two units:
			(c) $\langle x_1 \rangle$,  
			(d) $\langle x_2 \rangle$,  
			(e) $\langle p_1 \rangle$,
			and  (f) $\langle p_2 \rangle$. 
			(g) Time evolution of the negativity ${\cal N}$.
			The parameters are
			$\Delta = -0.6, \gamma_{1} = 0.4, \gamma_{2} = 0.1, 
			\theta = \pi$, $\eta = 0.3$, 
			$D_h = -0.99$, $D_c = 1$ ($D_x = 0.005$ and $D_p = 0.995$),
			and $\phi_j = 0$ and $\kappa_j = 0.25$ 
			for both $j = 1,2$.
			In (f), the black line represents the value for the steady state of the system without performing measurement.
		}
		\label{fig_7}
	\end{center}
\end{figure}

Figure~\ref{fig_7} shows the behavior of the system under continuous measurement in the semiclassical regime.
The parameters are the same as in Figs.~\ref{fig_3}(b,~d,~f), namely, the uniform state of the system has been destabilized by the Turing instability.
Considering that the nonuniformity is more pronounced in the position variable $x$ than in the momentum variable $p$ in Fig.~\ref{fig_3}(d), we set $\phi_j  = 0$ and perform the measurement on the quadrature $x_j = (a_j + a_j^\dag)/2$ ($j=1, 2$), which is conjugate to the momentum $p_j$, of both units.
We set the measurement efficiency as $\kappa_j = 0.25$ ($j = 1, 2$) for both units and the initial state of the whole system as the two-mode vacuum state.

\begin{figure} [htbp]
	\begin{center}
		\includegraphics[width=0.95\hsize,clip]{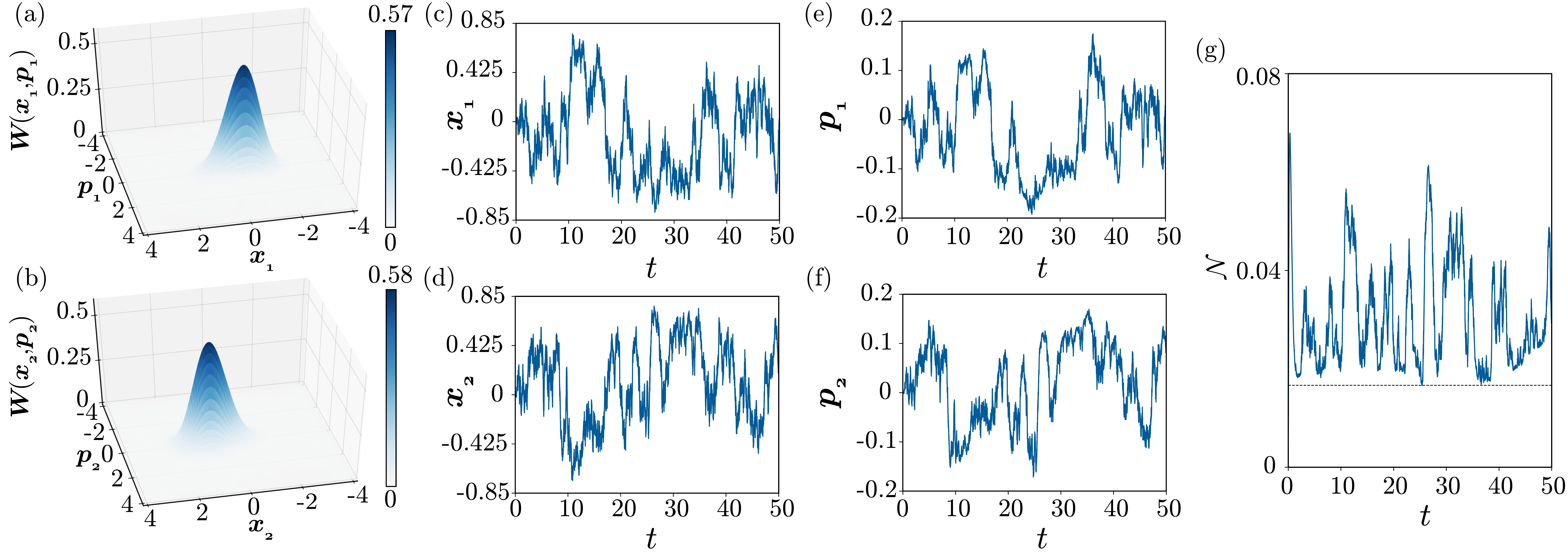}
		\caption{
			\textbf{Turing instability under continuous quantum measurement in the weak quantum regime.}
			(a, b) 3D snapshot plots of the Wigner distributions $W(x_1, p_1)$ and $W(x_2, p_2)$ at $t = 49.3 $.
			(c, d, e, f) Time evolution of the average values of the position and momentum 
			operators for two units:
			(c) $\langle x_1 \rangle$,  
			(d) $\langle x_2 \rangle$,  
			(e) $\langle p_1 \rangle$, 
			and (f) $\langle p_2 \rangle$. 
			(g) Time evolution of the negativity ${\cal N}$.
			The parameters are
			$\Delta = -0.6, \gamma_{1} = 1.2, \gamma_{2} = 0.5, 
			\theta = \pi$, $\eta = 0.3$, 
			$D_h = -0.99$, $D_c = 1$ ($D_x = 0.005$ and $D_p = 0.995$),
			and $\phi_j = 0$ and $\kappa_j = 0.25$ for both $j = 1,2$.
			In (f), the black line represents the value for the steady state of the system without performing measurement.
		}
		\label{fig_8}
	\end{center}
\end{figure}

\begin{figure} [htbp]
	\begin{center}
		\includegraphics[width=0.95\hsize,clip]{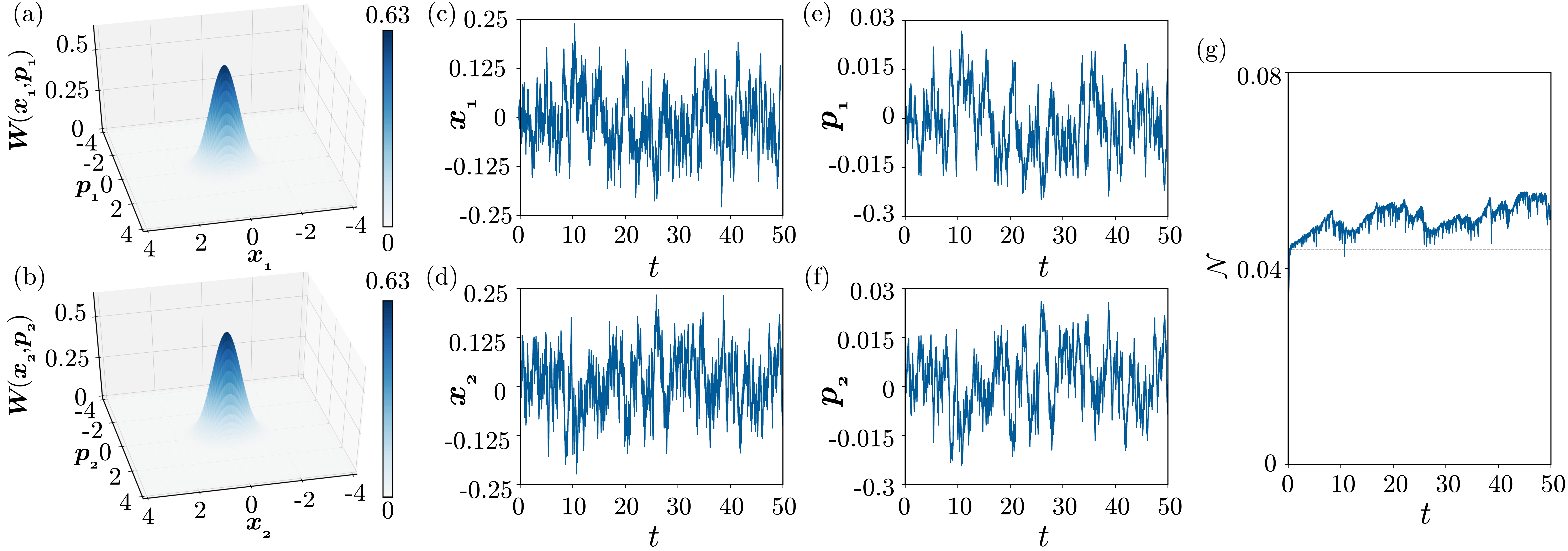}
		\caption{
			\textbf{Turing instability under continuous quantum measurement in the strong quantum regime.}
			(a, b) 3D snapshot plots of the Wigner distributions $W(x_1, p_1)$ and $W(x_2, p_2)$ at $t = 50$.
			(c, d, e, f) Time evolution of the average values of the position and momentum 
			operators for two units:
			(c) $\langle x_1 \rangle$,  
			(d) $\langle x_2 \rangle$,  
			(e) $\langle p_1 \rangle$, 
			and (f) $\langle p_2 \rangle$. 
			(g) Time evolution of the negativity ${\cal N}$.
			The parameters are
			$\Delta = -0.6, \gamma_{1} = 6.2, \gamma_{2} = 3, 
			\theta = \pi$, $\eta = 0.3$, 
			$D_h = -0.99$, $D_c = 1$ ($D_x = 0.005$ and $D_p = 0.995$),
			and $\phi_j = 0$ and $\kappa_j = 0.25$ for both $j = 1,2$.
			In (f), the black line represents the value for the steady state of the system without performing measurement.
		}
		\label{fig_9}
	\end{center}
\end{figure}

Figures~\ref{fig_7}(a) and (b) show the instantaneous marginal Wigner distributions 
$W(x_1, p_1)$ of $\rho_1$ and $W(x_2, p_2)$ of $\rho_2$ at time $t = 50$ sufficiently after the initial transient, obtained by a DNS of SME~(\ref{eq:2s_sme}). 
In contrast to Fig.~\ref{fig_3}(f), these Wigner distributions are not stationary and continue to fluctuate due to the continuous measurement. 
Each distribution is localized around either of the two stable fixed points of classical system (\ref{eq:2s_ctraj}) and tends to take the opposite state from the other one.

The anticorrelation between the states of the two units is evident in Figs.~\ref{fig_7}(c-f), where the time evolution of the average values of the position and momentum operators of both units,
$\mean{x_j} = \Tr[( (a_j + a^\dag_j)/2 )\rho ]$ and
$\mean{p_j} = -i \Tr[( (a_j - a^\dag_j)/2 )\rho ]$ ($j = 1,2$),
obtained from a single stochastic trajectory of quantum SME (\ref{eq:2s_sme}) are plotted.
The two units randomly alternate between the two nonuniform states and tend to take opposite states from each other.
This clearly indicates that the symmetry preserved by quantum noise is broken and that the asymmetry caused by the Turing instability in the classical sense is revealed by the extraction of information on the $x$ variables of the two units via continuous measurement.

Figure~\ref{fig_7}(g) shows the time evolution of the negativity $\cal{N}$ under the continuous measurement. The two units are clearly entangled and the degree of entanglement  continuously fluctuates around the value of $\cal{N}$ in the steady state when the measurement is not performed.

Similarly, Fig.~\ref{fig_8} shows the effect of continuous measurement in the weak quantum regime shown in Fig.~\ref{fig_4}. We observe qualitatively similar results to those for the semiclassical case in Fig.~\ref{fig_7} in the quantum regime.
Although the nonuniformity is less pronounced, the negativity is slightly larger on average,  and the fluctuations are stronger due to the effect of the stronger quantum measurement noise.
Notably, the negativity takes larger values than the case without performing measurement, indicating that the symmetry breaking due to the continuous measurement induces stronger entanglement in this regime.

Finally, we show in Fig.~\ref{fig_9} the effect of continuous measurement in the strong quantum regime shown in Fig.~\ref{fig_5}. 
Although the fluctuations are stronger due to the effect of the stronger quantum measurement noise than the two previous cases, the nonuniformity between two single units, which was quite small  in Fig.~\ref{fig_5}, is enhanced and more explicitly observed under the continuous measurement. Additionally, the negativity takes larger values than the case without measurement also in this strong quantum regime.
See also the Supplementary Movies for the time evolution of the marginal Wigner distributions of the two units.

\section{Concluding remarks}

We have theoretically demonstrated that Turing instability can occur in a quantum dissipative system. We showed that a degenerate parametric oscillator with nonlinear damping can be regarded as a quantum activator-inhibitor unit and that diffusive coupling between two such quantum activator-inhibitor units can give rise to Turing instability when the diffusivities of the activator and inhibitor variables are appropriately chosen.
Due to the Turing instability, the system becomes nonuniform but still remains in a symmetrically mixed state by the effect of  quantum noise. Further performing continuous quantum measurement breaks the symmetry and reveals the asymmetry between the two units.

We suppose that the physical setup assumed in our model can, in principle, be implemented by using currently available experimental devices.
The quantum activator-inhibitor unit is essentially a degenerate parametric oscillator with nonlinear damping~\cite{tezak2017low}.
The coupling terms via squeezing can be implemented by adjusting the single-mode squeezing parameter of the two quantum activator-inhibitor systems and introducing two-mode squeezing~\cite{nurdin2017linear}.
The dissipative coupling term could be realized by indirectly coupling the two oscillators through an additional cavity and adiabatically eliminating it~\cite{yang2017anti};
similar approaches have also been proposed for realizing dissipative couplings between ensembles of atoms~\cite{xu2014synchronization} and optomechanical Stuart-Landau oscillators~\cite{walter2015quantum}.
Another possible approach to the experimental realization of the proposed setups would be to use ``membrane-in-the-middle'' optomechanics~\cite{thompson2008strong}.
Physical implementations of single-mode squeezing and nonlinear damping~\cite{nunnenkamp2010cooling}, dissipative coupling~\cite{walter2015quantum}, and two-mode squeezing~\cite{tan2013dissipation} have also been proposed.
We expect that our numerical results for the Wigner distributions can be experimentally observed via quantum tomography~\cite{leonhardt1997measuring}. 
The experimental implementation of the continuous quantum measurement has also been reported recently~\cite{minev2019catch}.

In this study, we numerically analyzed a pair of quantum activator-inhibitor units that exhibits Turing instability in the classical, deterministic limit. 
For classical systems, analytical perturbative approaches have been applied to the classical master equation for predicting stochastic Turing patterns~\cite{biancalani2010stochastic, biancalani2011stochastic, asslani2012stochastic, asllani2013linear}.
We may be able to employ similar perturbative approaches for the quantum master equation~\cite{lee2014entanglement} and analyze the quantum Turing instability in more detail.	

The quantum activator-inhibitor unit could also be implemented by using quantum spin systems, which is interesting because small quantum spin systems may help us cope with the exponential increase in the
dimensions of the Hilbert space for large quantum networks~\cite{roulet2018synchronizing}. 
Similar to previous studies that discussed the Kerr effects~\cite{lorch2016genuine, amitai2018quantum} and quantum jumps~\cite{lee2012collective} 
in nonequilibrium pattern formation in quantum dissipative systems, clarifying the relationship between the Turing instability and strong quantum effects would be important.
A more detailed systematic  analysis on the relationship between Turing instability and entanglement is also a future study.

Although we analyzed only the minimal two-unit setup in this study, 
we may further consider Turing instability in larger networks of quantum activator-inhibitor units,
similar to the Turing instability in networks of classical activator-inhibitor systems~
\cite{othmer1971instability, othmer1974non, nakao2010turing, petit2017theory, muolo2019patterns}.
Compared to previous studies on quantum effects on nonlinear optical pattern formation~\cite{lugiato1992quantum, gatti1995quantum}, which are not easy to analyze even numerically because calculations of all operator products are required~\cite{zambrini2002macroscopic},
the activator-inhibitor system proposed in this study can be extended to larger networks more easily.
Thus, it may be used to reveal the novel emergence of self-organized patterns in quantum dissipative systems, similar to previous studies on the Kuramoto transition~\cite{lee2014entanglement}, quantum chimera states~\cite{bastidas2015quantum}, and oscillation death~\cite{ishibashi2017oscillation} in globally connected quantum Stuart-Landau oscillator networks. 
Though we focused on a pair of coupled activator-inhibitor units in this study, we may also be able to further couple many units on a lattice or network of units and analyze the spatio-temporal pattern formation in fully quantum mechanical dissipative systems.

The quantum Turing instability may also find technical applications.
For example, signal amplification near bifurcation points has been theoretically investigated in classical biological systems~\cite{mora2011biological, munoz2018colloquium} 
and other classical~\cite{wiesenfeld1986small}, nanoscale \cite{buks2006mass}, and quantum~\cite{dutta2019critical} nonlinear systems,
and signal amplifiers using nonlinear bifurcation have been experimentally implemented~\cite{siddiqi2004rf}. 
Similarly, the Turing bifurcation in quantum dissipative systems may also offer new engineering applications for quantum signal amplification and quantum sensing. 

As Turing instability is a paradigm of nonequilibrium self-organization in classical systems \cite{reinitz2012pattern}, we believe that our results on the possibility of Turing instability in quantum dissipative systems 
also play an essentially important role in studying
self-organization in quantum systems 
and will be relevant in the growing field of quantum technology.

\section{Acknowledgments.}
Numerical simulations were performed by using the QuTiP numerical
toolbox~\cite{johansson2012qutip,johansson2013qutip}. 
We acknowledge 
JSPS KAKENHI JP17H03279, JP18H03287, JPJSBP120202201, JP20J13778,  JP22K14274, JP22K11919, JP22H00516 and 
JST CREST JP-MJCR1913 for financial support. 
\section{Methods}
\subsection{Classical activator-inhibitor systems and Turing instability}

A classical activator-inhibitor system is generally described by
\begin{align}
	\label{eq:gen1s}
	\dot{x}  & = f(x, p),
	\cr
	\dot{p}  & = g(x, p),
\end{align}
where $(\dot{})$ denotes the time derivative and $x$ and $p$ represent the activator and inhibitor variables, respectively.
We assume that this system has a stable fixed point at $(x, p) = (\bar{x}, \bar{p})$. 
Denoting small variations from $(\bar{x}, \bar{p})$ as $\delta x = x - \bar{x}$ and $\delta p = p - \bar{p}$ and linearizing Eq.~(\ref{eq:gen1s}), we obtain
\begin{align}
	\label{eq:gen1s_lin}
	\frac{d}{dt}
	\left( \begin{matrix}
		{\delta x}  \\
		{\delta p}  \\
	\end{matrix} \right)
	&=
	\left( \begin{matrix}
		f_x
		& 
		f_p
		\\
		g_x
		&
		g_p
		\\
	\end{matrix} \right)
	\left( \begin{matrix}
		\delta x  \\
		\delta p  \\
	\end{matrix} \right),
\end{align}
where we assume that the coefficients satisfy
\begin{align}
	f_x = \partial f/\partial x|_{(\bar{x}, \bar{p})} > 0,
	\quad
	f_p = \partial f/\partial p|_{(\bar{x}, \bar{p})} < 0,
	\cr
	g_x = \partial g/\partial x|_{(\bar{x}, \bar{p})} > 0,
	\quad
	g_p = \partial g/\partial p|_{(\bar{x}, \bar{p})} < 0.
	\label{eq:aiconditions}
\end{align}
These are the conditions in which $x$ is the activator and $p$ is the inhibitor.
These standard conditions can be eased in more general settings~\cite{madzvamuse2010stability}, but we restrict our focus on the cases satisfying these conditions. 

We consider two diffusively coupled activator-inhibitor units with identical properties, described by
\begin{align}
	\label{eq:gen2s}
	\left( \begin{matrix}
		\dot{x}_1  \\
		\dot{p}_1  \\
		\dot{x}_2  \\
		\dot{p}_2  \\
	\end{matrix} \right)=
	\left( \begin{matrix}
		f(x_1, p_1) + D_x (x_2 - x_1) \\
		g(x_1, p_1) + D_p (p_2 - p_1) \\
		f(x_2, p_2) + D_x (x_1 - x_2) \\
		g(x_2, p_2) + D_p (p_1 - p_2) \\
	\end{matrix} \right),
\end{align}
where $D_x$ and $D_p$ represent the diffusion constants of the activator and inhibitor variables, respectively.
This coupled system has a trivial fixed point $(x_1, p_1, x_2, p_2) = (\bar{x}, \bar{p}, \bar{x}, \bar{p})$, which corresponds to a uniform state of the whole system. 

In Turing instability, contrary to our intuition, this uniform state can be destabilized by the effect of diffusion when the parameters satisfy appropriate conditions.
To see this, we linearize Eq.~(\ref{eq:gen2s}) as
\begin{align}
	\label{eq:gen2s_lin}
	\frac{d}{dt}
	\left( \begin{matrix}
		\delta{x}_1  \\
		\delta{p}_1  \\
		\delta{x}_2  \\
		\delta{p}_2  \\
	\end{matrix} \right)=
	\left( \begin{matrix}
		f_x - D_x & g_x & D_x & 0
		\\
		f_p & g_p - D_p & 0 & D_p
		\\
		D_x & 0 & f_x - D_x & g_x 
		\\
		0 & D_p & f_p & g_p - D_p
		\\
	\end{matrix} \right)
	&\left( \begin{matrix}
		\delta x_1  \\
		\delta p_1  \\
		\delta x_2  \\
		\delta p_2  \\
	\end{matrix} \right),
\end{align}
where $\delta x_j = x_j - \bar{x}$ and $\delta p_j = p_j - \bar{p}$ ($j=1, 2$) are small variations.
The maximum eigenvalue of the Jacobian matrix in Eq.~(\ref{eq:gen2s_lin}) is given by
\begin{align}
	\lambda_{max} = - (D_x + D_p) + \frac{f_x + g_p}{2} 
	+ \sqrt{ (D_p - D_x)(D_p - D_x + f_x - g_p) + \frac{ (f_x - g_p)^2}{4} + f_p g_x}.
\end{align}
Therefore, when $\lambda_{max} > 0$, namely, when
\begin{align}
	4 D_x D_p - 2 D_p f_x - 2 D_x g_p + f_x g_p - f_p g_x < 0,
	\label{eq:turingcondition}
\end{align}
the uniform fixed point $(x_1, p_1, x_2, p_2) = (\bar{x}, \bar{p}, \bar{x}, \bar{p})$ of the coupled system destabilizes.

In our model, the functions $f$ and $g$ are given by
\begin{align}
	f(x,p) &= \frac{ 2 \gamma_{2} - \gamma_{1}}{2} x + \Delta  p  
	- \gamma_{2} x  (x^{2} + p^{2}) 
	- 2 \eta ( x \cos \theta + p \sin \theta), 
	\cr
	g(x,p) &=   - \Delta  x + \frac{ 2 \gamma_{2} - \gamma_{1}}{2} p  
	- \gamma_{2} p  (x^{2} + p^{2}) 
	+ 2  \eta ( - x \sin \theta + p \cos \theta ),
\end{align}
where $\gamma_1, \gamma_2, \eta$, and $\Delta$ are parameters.
The derivatives of $f$ and $g$ at this fixed point are given by
\begin{align}
	f_x = \frac{2 \gamma_{2} - \gamma_{1}}{2} - 2 \eta \cos \theta,
	\quad
	f_p = \Delta - 2 \eta \sin \theta,
	\cr
	g_x = -\Delta - 2 \eta \sin \theta,
	\quad
	g_p = \frac{2 \gamma_{2} - \gamma_{1}}{2}
	+ 2  \eta \cos \theta.
\end{align}
With the parameter values used in the present study, 
the single system in Eq.~(\ref{eq:gen1s}) has a stable fixed point at $(x, p) = (\bar{x}, \bar{p}) = (0, 0)$,
the conditions in Eq.~(\ref{eq:aiconditions}) for the single system to be of the activator-inhibitor type are satisfied, and the condition for the Turing instability in Eq.~(\ref{eq:turingcondition}) can be satisfied 
for a pair of diffusively coupled quantum activator-inhibitor units.

As the Turing instability takes place, the trivial fixed point $(0, 0, 0, 0)$ of the system is destabilized, and two new stable fixed points,
\begin{align}
	(x_1, p_1, x_2, p_2) = (A, B, -A, -B),\ (-A, -B, A, B),
\end{align}
which correspond to the nonuniform states of the whole system, arise via the supercritical pitchfork bifurcation, where
\begin{align}
	A &= R \cos \Theta, 
	\cr
	B &= R \sin \Theta,
	\cr
	R &= \sqrt{\frac{1}{\gamma_2} \left( \frac{2\gamma_2 - \gamma_3}{2} - (D_p + D_x) 
		+ \sqrt{4 \eta^2 - 4\eta \cos \theta(D_p - D_x) + (D_p - D_x)^2 - \Delta^2} \right)},
	\cr 
	\Theta &= \frac{1}{2} \Big( \pi + \arctan \left( \frac{2\eta \sin \theta}{ 2\eta \cos \theta - (D_p - D_x) } \right)
	- \sin^{-1} \frac{\Delta}{\sqrt{4 \eta^2 - 4\eta \cos\theta (D_p - D_x) + (D_p - D_x)^2}} \Big).
	\cr
\end{align}

With the parameter values used in the Results section, the derivatives of $f$ and $g$ are $f_x = 0.5$, $f_p = -0.6$, $g_x = 0.6$, and $g_p = -0.7$. 
In Figs.~\ref{fig_3} and \ref{fig_4}, the maximum eigenvalue of the uniform fixed point is $\lambda_{max} \approx 0.3724 > 0$; hence, Turing instability has already occurred.

\subsection{Quantum-classical correspondence via the Wigner distribution}

We generally consider a quantum dissipative system with $N$ modes, which is coupled with $n$ reservoirs. We denote by $a_1, ..., a_N$ and $a_1^{\dag}, ..., a_N^{\dag}$ the annihilation and creation operators of the system, respectively. %
A general form of the QME describing this quantum dissipative system is given by
\begin{align}
	\label{eq:gen_me}
	\dot{\rho}
	= 
	-i[H, \rho] 
	+ \sum_{j=1}^{n} \mathcal{D}[L_{j}]\rho,
\end{align}
where $\rho$ is the density operator representing the system state, $H$ is a system Hamiltonian, $L_{j}$ is a coupling operator between the system and $j$th reservoir $(j=1,\ldots,n)$, and $\mathcal{D}[L]\rho = L \rho L^{\dag} - (\rho L^{\dag} L + L^{\dag} L \rho)/2$ is the Lindblad form~\cite{gardiner1991quantum,carmichael2007statistical}. 

By using the standard method of phase-space representation~\cite{gardiner1991quantum,carmichael2007statistical},
we can introduce the Wigner distribution $W({\bm \alpha}) \in {\mathbb R}$ of $\rho$ as
\begin{align}
	W({\bm \alpha}) =
	\frac{1}{\pi^{2N}} 
	\int 
	\exp \left( \sum_{j} (-\lambda_j \alpha_j^{*}+\lambda_j^{*} \alpha_j) \right)\Tr \left\{\rho  D( \bm{\lambda}, \bm{a}) \right\}
	d^{2N} \bm{\lambda},
\end{align}
where ${\bm \alpha} = ( \alpha_1, \alpha^*_1, \ldots, \alpha_N, \alpha^*_N ) \in {\mathbb C}^{2N}$
represents the state variable in the $2N$-dimensional phase space,
$D( \bm{\lambda}, \bm{a}) = \exp \left( \sum_{j} (\lambda_j a_j^{\dagger} - \lambda_j^{*} a_j) \right)$, 
$d^{2N} \bm{\lambda} = d \lambda_1 d \lambda^*_1 \ldots d \lambda_N d \lambda^*_N$, 
$\alpha_j, \alpha_j^* \in \mathbb{C}$, $\lambda_j,\lambda_j^* \in \mathbb{C}$,
and ${}^*$ indicates complex conjugate.
QME~(\ref{eq:gen_me}) for the density operator $\rho$ can be transformed into a partial differential equation for the Wigner distribution $W({\bm \alpha})$~\cite{gardiner1991quantum,carmichael2007statistical}, given by
\begin{align}
	\label{eq:gen_pde}
	\frac{\partial}{\partial t} W({\bm \alpha}) = \mathcal{L}_p W({\bm \alpha}).
\end{align}
Here, the differential operator $\mathcal{L}_p$ can be explicitly calculated from Eq.~(\ref{eq:gen_me}) by using the standard calculus~\cite{gardiner1991quantum, carmichael2007statistical}.

When the quantum effect is relatively weak, we may neglect the derivative terms higher than the second order in Eq.~(\ref{eq:gen_pde}). 
Then, by introducing a real-valued representation of the phase-space variable, $\bm{X} = (x_1, p_1, \ldots, x_N, p_N)$ with $\alpha_j = x_j + i p_j$ ($j=1, ...,N$), 
we can approximate Eq.~(\ref{eq:gen_pde}) by the semiclassical FPE for $W({\bm X})$, 
\begin{align}
	\label{eq:gen_fpe}
	\frac{\partial}{\partial t} W({\bm X}) = 
	\left( -\frac{\pa}{\pa {\bm X}}{\bm A}({\bm X}) + \frac{1}{2}\frac{\pa^2}{\pa {\bm X}^2}
	{\bm D}({\bm X})  \right) W({\bm X}).
\end{align}
Here, ${\bm A}(\bm{X}) \in \mathbb{R}^{2N}$ is the the drift vector, and ${\bm D}(\bm{X}) \in \mathbb{R}^{2N \times 2N}$ represents the diffusion matrix.
The SDE corresponding to the above FPE is given by
\begin{align}
	\label{eq:gen_sde}
	d\bm{X} = {\bm A}(\bm{X}) dt +  {\bm G}(\bm{X})d\bm{W}.
\end{align}
Here, ${\bm A}({\bm X})$ is the same as in Eq.~(\ref{eq:gen_fpe}), the matrix ${\bm G}(\bm{X}) \in {\mathbb R}^{2N}$ represents the noise intensity satisfying ${\bm G}(\bm{X}) {\bm G}^T(\bm{X}) = {\bm D}(\bm{X})$ with $T$ representing the matrix transpose, and $d\bm{W} = ( dw_1, \ldots, dw_{2N}) \in {\mathbb R}^{2N}$ represents a vector of independent Wiener processes satisfying $\mean{dw_k(t)dw_l(t)} = \delta_{kl} dt$ with $k,l = 1, \ldots, 2N$. 
The deterministic trajectory in the classical limit is given by the deterministic term of the SDE, namely, $\dot{\bm{X}} = {\bm A}(\bm{X})$.

\subsection{Derivation of Semiclassical Fokker-Planck and stochastic differential equations}

We here give explicit forms of the approximate Fokker-Planck equation (FPE) and semiclassical stochastic differential equation (SDE) derived from quantum master equation (QME)~(3) in the Results section for two diffusively coupled quantum activator-inhibitor units,
\begin{align}
	\label{eq:s1_me}
	\dot{\rho}
	& = 
	\sum_{j=1,2} \left( -i 
	\left[\Delta a_{j}^{\dag}a_{j}
	+ i \eta ( a_{j}^2 e^{-i \theta} - a_{j}^{\dag 2} e^{ i \theta})
	, \rho \right]
	+ \gamma_{1}\mathcal{D}[a_{j}]\rho 
	+ \gamma_{2}\mathcal{D}[a_{j}^{2}]\rho 
	\right)
	\cr
	& - i \left[  i \frac{D_{h}}{4}\{ (a_1 - a_2)^2 - (a_1^\dag - a_2^\dag)^2 \}, \rho \right]
	+ 
	D_{c} \mathcal{D}[a_{1} - a_{2}]\rho.
\end{align}
By using the standard calculus for the phase-space representation~\cite{gardiner1991quantum, carmichael2007statistical}, we can derive the following partial differential equation representing the time evolution of the Wigner distribution 
$W({\bm \alpha}, t)$ for $\bm{\alpha} = (\alpha_1, \alpha_1^*,\alpha_2,\alpha_2^*)$ from Eq.~(\ref{eq:s1_me}) : 
\begin{align}
	\label{eq:s1_tewig}
	\frac{\pa W(\bm{\alpha}, t)}{\pa t} 
	=& \sum_{j=1}^{2} \Big[  
	-\Big( \frac{\partial}{\partial \alpha_j} A_{\alpha_j} + c.c. \Big) 
	+ \frac{1}{2} 
	\Big(
	\frac{\partial^2}{\partial \alpha_j \partial \alpha^*_j}D_{\alpha_j, \alpha^*_j} 
	+ 
	\frac{\partial^2}{\partial \alpha_j \partial \alpha^*_{\overline{j}}}D_{\alpha_j, \alpha^*_{\overline{j}}} 
	+
	c.c. 
	\Big) 
	\cr
	&+
	\Big(
	\frac{\gamma_{2}}{4}
	\frac{\pa^3}{\pa^2 \alpha_j \pa \alpha_j^{*}} \alpha_j 
	+
	c.c. 
	\Big)
	\Big]W(\bm{\alpha}, t), 
\end{align}
where
\begin{align}
	&A_{\alpha_j} =
	\left(\frac{2 \gamma_{2} - \gamma_1}{2} - i \Delta \right) \alpha_j  
	- \gamma_{2} \alpha_j^{*} \alpha_j^{2} 
	- 2 \eta e^{i \theta}\alpha_j^{*} 
	+  
	\frac{D_{c}}{2}(\alpha_{\overline{j}} - \alpha_j) + 
	\frac{D_{h}}{2}(\alpha^*_{\overline{j}} - \alpha^*_j), 
	\cr
	&D_{\alpha_j, \alpha^*_j} = \frac{\gamma_1 + D_{c}}{2} + 2 \gamma_{2} \left( \abs{\alpha_j}^2 -  \frac{1}{2} \right),
	\quad
	D_{\alpha_j, \alpha^*_{\overline{j}}} = - \frac{D_{c}}{2}.
\end{align}
Here and henceforth, $\overline{j}$ denotes $\overline{j} = 2$ when $j=1$ and $\overline{j}=1$ when $j = 2$, and $c.c.$ denotes the complex conjugate. 

In the semiclassical regime where $\gamma_2$ is sufficiently small,
the third-order derivative terms in Eq.~(\ref{eq:s1_tewig}) can be neglected 
\cite{lee2013quantum,lorch2016genuine, ishibashi2017oscillation} and 
the coefficients of the second-order derivative terms are positive.
Therefore, Eq.~(\ref{eq:s1_tewig}) can be approximated by the FPE 
\begin{align}
	\label{eq:s1_fpe}
	&\frac{\pa W(\bm{\alpha}, t)}{\pa t} 
	= \sum_{j=1}^{2} \Big[  
	-\Big( \frac{\partial}{\partial \alpha_j} A_{\alpha_j} + c.c. \Big) 
	+ \frac{1}{2} 
	\Big(
	\frac{\partial^2}{\partial \alpha_j \partial \alpha^*_j}D_{\alpha_j, \alpha^*_j} 
	+ 
	\frac{\partial^2}{\partial \alpha_j \partial \alpha^*_{\overline{j}}}D_{\alpha_j, \alpha^*_{\overline{j}}} 
	+
	c.c. 
	\Big)
	\Big]W(\bm{\alpha}, t).
\end{align}
Using a real-valued representation, i.e., $\bm{X} = (x_1, p_1, x_2, p_2)$ with 
$\alpha_j = x_j + i p_j~(j = 1,2)$, Eq.~(\ref{eq:s1_fpe}) can be rewritten as
\begin{align}
	\label{eq:s1_fpe_re}
	&\frac{\pa W(\bm{X}, t)}{\pa t} 
	= \sum_{j=1}^{2} \Big[  
	-\Big(  \frac{\partial}{\partial x_j} A_{x_j} + \frac{\partial}{\partial p_j} A_{p_j} \Big) 
	\cr
	& + \frac{1}{2} 
	\Big(
	\frac{\partial^2}{\partial x_j \partial x_j} D_{x_j, x_j} 
	+ 
	\frac{\partial^2}{\partial p_j \partial p_j} D_{p_j, p_j} 
	+ 
	\frac{\partial^2}{\partial x_j \partial x_{\overline{j}}} D_{x_j, x_{\overline{j}}}
	+ 
	\frac{\partial^2}{\partial p_j \partial p_{\overline{j}}} D_{p_j, p_{\overline{j}}}
	\Big) 
	\Big]W(\bm{X}, t), 
\end{align}
where
\begin{align}
	&A_{x_j} = 
	\frac{2 \gamma_{2} - \gamma_{1}}{2} x_j  + \Delta  p_j  
	- \gamma_{2} x_j  (x_j^{2} + p_j^{2}) 
	- 2 \eta ( x_j \cos \theta + p_j \sin \theta)  
	+ D_x(x_{\overline{j}} - x_j),
	\cr
	&A_{p_j} = 
	- \Delta  x_j + \frac{2 \gamma_{2} - \gamma_{1}}{2} p_j    
	- \gamma_{2} p_j  (x_j^{2} + p_j^{2}) 
	+ 2  \eta ( - x_j \sin \theta + p_j \cos \theta ) 
	+ D_{p}(p_{\overline{j}} - p_j),
	\cr
	&D_{x_j, x_j} =  D_{p_j, p_j} = \frac{\gamma_1 + D_{c}}{4} + \gamma_{2} \left( x_j^2 + p_j^2  -  \frac{1}{2} \right),
	\cr
	&D_{x_j, x_{\overline{j}}} = D_{p_j, p_{\overline{j}}} = - \frac{D_{c}}{4}. 
\end{align}
Thus, the drift vector is given by 
$\bm{A}(\bm{X}) = (A_{x_1}, A_{p_1}, A_{x_2}, A_{p_2}) $
and the diffusion matrix ${\bm D}(\bm{X})$ is expressed as
\begin{align}
	{\bm D}(\bm{X})
	&=\frac{1}{2}\left(\begin{array}{cccc}
		v_{1} & 0 & -D_c / 2 & 0 \\
		0 & v_{1} & 0 & -D_c / 2 \\
		-D_c / 2 & 0 & v_{2} & 0 \\
		0 & -D_c / 2 & 0 & v_{2}
	\end{array}\right),
\end{align}
where we defined
\begin{align}
	v_{j}= \frac{1}{2}(\gamma_1 + D_{c}) + 2 \gamma_{2} \left( x_j^2 + p_j^2  -  \frac{1}{2} \right).
\end{align}

The SDE corresponding to FPE (\ref{eq:s1_fpe_re}) is given by
\begin{align}
	\label{eq:semildv}
	d\bm{X}(t) = {\bm A}(\bm{X}(t)) dt +  {\bm G}(\bm{X}(t))d\bm{W}(t),
\end{align}
where ${\bm G}({\bm X})$ satisfies ${\bm G}(\bm{X}) {\bm G}^T(\bm{X}) = {\bm D}(\bm{X})$  
and $d\bm{W}(t)$ $= ( dw_1(t),$ $dw_2(t),$  $dw_3(t)$, $dw_4(t))^T$ is a vector of independent Wiener processes satisfying $\mean{dw_k(t)dw_l(t)} = \delta_{kl} dt$ for $k,l = 1,2,3,4$. 

When $D_{c} = 0$, we have ${\bm G}(\bm{X}) = 
\operatorname{diag}\left( \sqrt{v_{1}/2}, \sqrt{v_{1}/2}, \sqrt{v_{2}/2}, \sqrt{v_{2}/2} \right)$. When $D_{c} \neq 0$, the diffusion matrix $\bm{D}(\bm{X})$ can be diagonalized by using the matrix
\begin{align}
	{\bm U}(\bm{X})
	&=\left(\begin{array}{cccc}
		0 & u_- & 0 & u_+ \\
		u_- & 0 & u_+ & 0 \\
		0 & 1 & 0 & 1 \\
		1 & 0 & 1 & 0
	\end{array}\right)
\end{align}
as
\begin{align}
	\bm{D}'(\bm{X})= \bm{U}^{-1}(\bm{X}) \bm{D}(\bm{X}) \bm{U}(\bm{X}) = 
	\operatorname{diag} \left( \Lambda_{-}, \Lambda_{-}, \Lambda_{+}, \Lambda_{+} \right),
\end{align}
where
\begin{align}
	u_{\pm}=-\frac{v_{1}-v_{2} \pm \sqrt{\left(v_{1}-v_{2}\right)^{2}+D_c^{2}}}{D_c}
\end{align}
and  
\begin{align}
	\Lambda_{\pm}=\frac{1}{4}\left(v_{1}+v_{2} \pm \sqrt{\left(v_{1}-v_{2}\right)^{2}+D_c^{2}}\right).
\end{align}
Thus, the matrix ${\bm G}(\bm{X})$ can be chosen as ${\bm G}(\bm{X}) = \bm{U}(\bm{X}) \sqrt{\bm{D}'(\bm{X})} \bm{U}^{-1}(\bm{X}) $~\cite{ishibashi2017oscillation}, i.e.,
\footnotesize
\begin{align}
	{\bm G}(\bm{X}) = \frac{1}{u_{+}-u_{-}}
	\left(
	\begin{array}{cccc}
		u_{+} \sqrt{\Lambda_{+}} - u_{-}\sqrt{\Lambda_{-}} & 0 & \sqrt{\Lambda_{+}}-\sqrt{\Lambda_{-}} & 0 \\
		0 & u_{+} \sqrt{\Lambda_{+}} - u_{-}\sqrt{\Lambda_{-}} & 0 & \sqrt{\Lambda_{+}}-\sqrt{\Lambda_{-}} \\
		\sqrt{\Lambda_{+}}-\sqrt{\Lambda_{-}} & 0 & u_{+} \sqrt{\Lambda_{-}} - u_{-}\sqrt{\Lambda_{+}} & 0 \\
		0 & \sqrt{\Lambda_{+}}-\sqrt{\Lambda_{-}} & 0 & u_{+} \sqrt{\Lambda_{-}} - u_{-}\sqrt{\Lambda_{+}}
	\end{array}
	\right).
\end{align}
\normalsize

\subsection{Direct numerical simulations of the quantum SDE}

In addition to the QME, we also perform direct numerical simulations of semiclassical SDE~(\ref{eq:semildv}) corresponding to FPE~(\ref{eq:s1_fpe_re})
to show the relationship of the distributions of the quantum states with the classical fixed points after the Turing instability.
For example, supplementary Figures~S1(a) and (b) show scatter plots of a stochastic trajectory of two diffusively coupled quantum activator-inhibitor units, and Figs.~S1(c) shows the 2D plot of the Wigner distribution 
$W(x_{1,2}, p_{1,2})$ in Fig.~3(f).
In Figs.~S1(a,~b), the states of units $1$ and $2$ stochastically go back and forth between the two stable fixed points due to quantum noise.
These scatter plots agree with the Wigner distributions distributed around the two stable fixed points in Fig~S1(c).

\subsection{Characterization of the quantum regime}

We characterize the degree of quantum effect as the nonlinear damping parameter $\gamma_2$ is varied by using the accuracy of the semiclassical approximation. 
The discrepancy between the semiclassical approximation and the original QME characterizes how deep the system is in the quantum regime. 
To keep the parameters of the corresponding classical systems unchanged, the linear damping parameter is chosen as $\gamma_1 = \gamma_1' + 2 \gamma_2$, where $\gamma_1'$ is a constant, and the other parameters are fixed to the same values as those used in the Results section.

\begin{figure} [htbp]
	\begin{center}
		\includegraphics[width=1\hsize,clip]{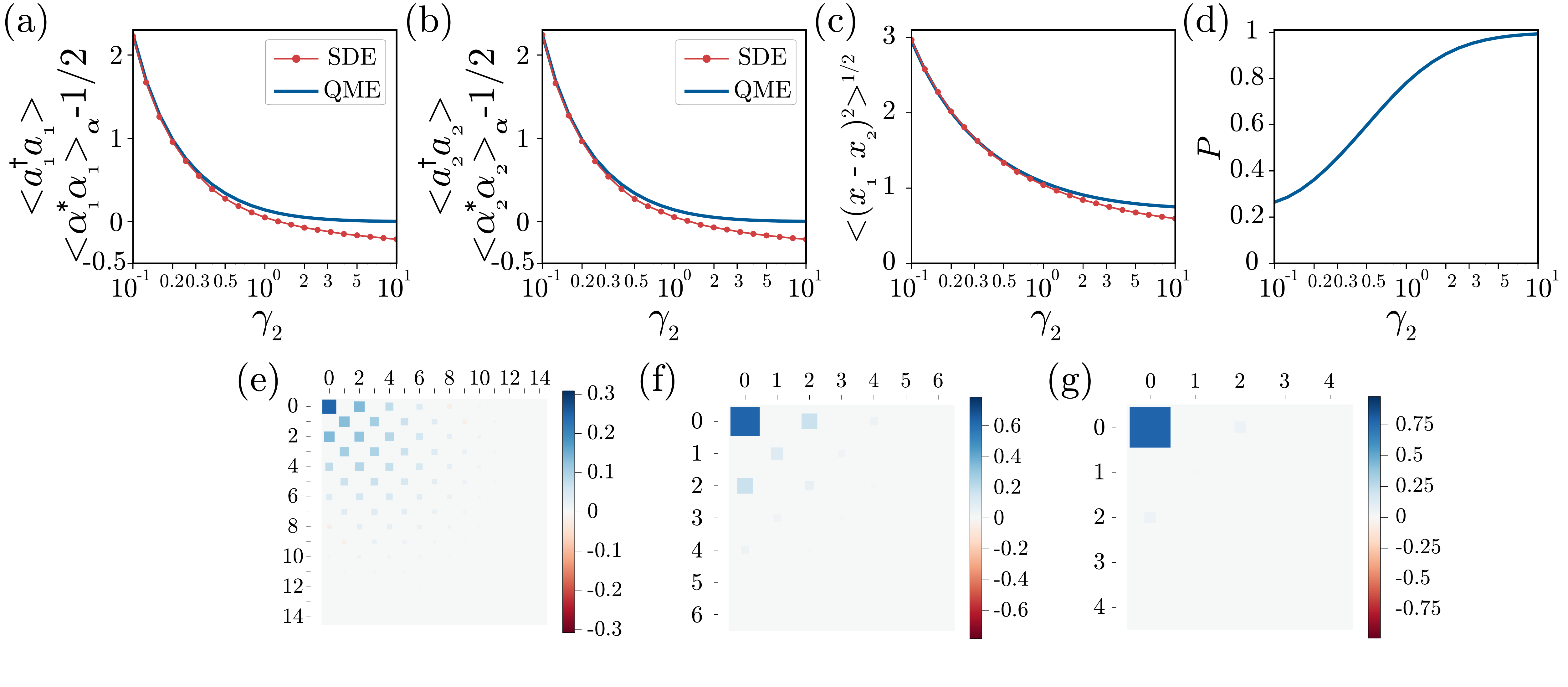}
		\caption{
			\textbf{Characterization of the quantum regime: average photon numbers,  nonuniformity, purity, and  elements of the density matrix of a single unit vs. $\bf{\gamma_2}$.} (a) Average photon number of unit $1$. (b) Average photon number of unit $2$.
			(c) Root mean squared distance $\sqrt{\mean{(x_1 - x_2)^2}}$ (d) Purity $P$. 
			(e-g) Elements of the density matrix of a single unit $\rho_1$ with respect to the number basis in the semiclassical (e), weak quantum (f), and strong quantum regime (g). In (a-c), results obtained from the semiclassical SDE $\mean{\alpha_j \alpha^*_j}_{\bm{\alpha}} - 1/2$ (red dots) and QME $\mean{a_j^\dag a_j}$ (blue lines) ($j=1, 2$) are shown, where $\mean{\alpha_j \alpha^*_j}_{\bm{\alpha}}$ is calculated as a time average of $\alpha_j(t) \alpha^*_j(t)$ over a time interval of length $30000$ after the initial transient.
			The parameters are $\Delta = -0.6, \theta = \pi$, $\eta = 0.3$, $D_h = -0.99$, $D_c = 1$ ($D_x = 0.005$ and $D_p = 0.995$),		
			and $\gamma_{1} = \gamma'_1 + 2 \gamma_2$ with $\gamma'_1 = 0.2$. In (e-g), $\gamma_2 = 0.1$ (e), $\gamma_2 = 0.5$ (f), $\gamma_2 = 3$ (g).
		}
		\label{fig_10}
	\end{center}
\end{figure}

\clearpage
Figures~\ref{fig_10}(a), (b) and (c)
plot the average numbers of photons in both units and the nonuniformity
$\sqrt{\mean{(x_1 - x_2)^2}}$ as functions of the nonlinear damping parameter $\gamma_2$.
Here, the average number of photons is calculated as an ensemble average $\mean{ a_j^\dag a_j } = \Tr[ a_j^\dag a_j \rho]~(j=1,2)$ of $a_j^\dag a_j$ obtained from the QME and as an average $\mean{\alpha_j \alpha^*_j}_{\bm{\alpha}}$ of $\alpha_j \alpha^*_j$ obtained from the semiclassical SDE, where the relation
\begin{align}
	\mean{\alpha_j \alpha^*_j}_{\bm{\alpha}} - 1/2
	\approx 
	\mean{a_j a_j^\dag + a_j^\dag a_j}/2 - 1/2
	= 
	\mean{a_j^\dag a_j}
\end{align}
holds approximately in the semiclassical regime. 
The semiclassical results well approximate the results of the QME in the regime with  small $\gamma_2$, and the error due to the semiclassical approximation gradually increases
with increasing $\gamma_2$.
Thus, when $\gamma_2 = 0.1$  (Figs. 2, 3, 6(c,~f), and 7), the semiclassical approximation is valid and the system is in the semiclassical regime, whereas when $\gamma_2 = 0.5$ (Figs. 4, 6(d,~g) and 8) and $\gamma_2 = 3$ (Figs. 5, 6(e,~h) and 9), the semiclassical approximation is no longer valid and the system is in the quantum regime. 
The degree of quantum effect can also be characterized by the purity as shown in Figs.~\ref{fig_10}(d), where the purity increases with the increase of $\gamma_2$. %
We also show in Figs.~\ref{fig_10}(e, f, g) the elements of the density matrix of a single unit $\rho_1$ with respect to the number basis in the semiclassical (e), weak quantum (f), and strong quantum regime (g).
We see that the energy level up to which the elements of the density matrix take non-zero value becomes lower and the discreteness of the energy spectrum becomes more prominent with the increase of $\gamma_2$.

\subsection{Negativity}
We use the negativity $\mathcal{N} = ({\left\|\rho^{\Gamma_{1}}\right\|_{1}-1})/{2}$ to quantify the quantum entanglement of the two units, where $\rho^{\Gamma_{1}}$ represents the partial transpose 
of the density operator $\rho$ of the two-mode system with units $1$ and $2$ with respect to unit $1$ and $\left\|X\right\|_{1}=\operatorname{Tr}|X|=\operatorname{Tr} \sqrt{X^{\dagger} X}$ 
\cite{zyczkowski1998volume, vidal2002computable}.
A non-zero negativity indicates that the two units are entangled.
{Note that the negativity ${\cal N}' =  ({\left\|\rho^{\Gamma_{2}}\right\|_{1}-1})/{2}$ calculated with respect to unit $2$ is equal to the negativity ${\cal N}$ calculated with respect to the unit $1$.}

\section*{Author contributions}
Both authors designed the study, carried out the analysis, and contributed to writing the paper. Y. K. performed numerical simulations.

\section*{Competing interests}
The authors declare no competing interests.

\section*{Data availability}
All data generated or analysed during this study are included in this published article and its supplementary information files.


\clearpage

\setcounter{page}{1}
\setcounter{section}{0}
\renewcommand{\thesection}{S\arabic{section}}
\setcounter{figure}{0}
\renewcommand{\figurename}{FIG. S}
\setcounter{equation}{0}
\renewcommand{\theequation}{S\arabic{equation}}

\begin{center}
	{\large \bf
		Turing instability in quantum activator-inhibitor systems  \\
		- Supplementary Information -  \\
		
	}
	{\bf Yuzuru Kato$^1$ and Hiroya Nakao$^2$} \\
	$^1$ Department of Complex and Intelligent Systems,
	Future University Hakodate, Hokkaido 041-8655, Japan 
	(Corresponding author: katoyuzu@fun.ac.jp) \\
	$^2$ Department of Systems and Control Engineering,
	Tokyo Institute of Technology, Tokyo 152-8552, Japan \\
	\date{\today}
	
\end{center}

\section*{Abstract}
A Supplementary Figure and three Supplementary Movies are available.  


\clearpage
\begin{figure} [htbp]
	\begin{center}
		\includegraphics[width=0.95\hsize,clip]{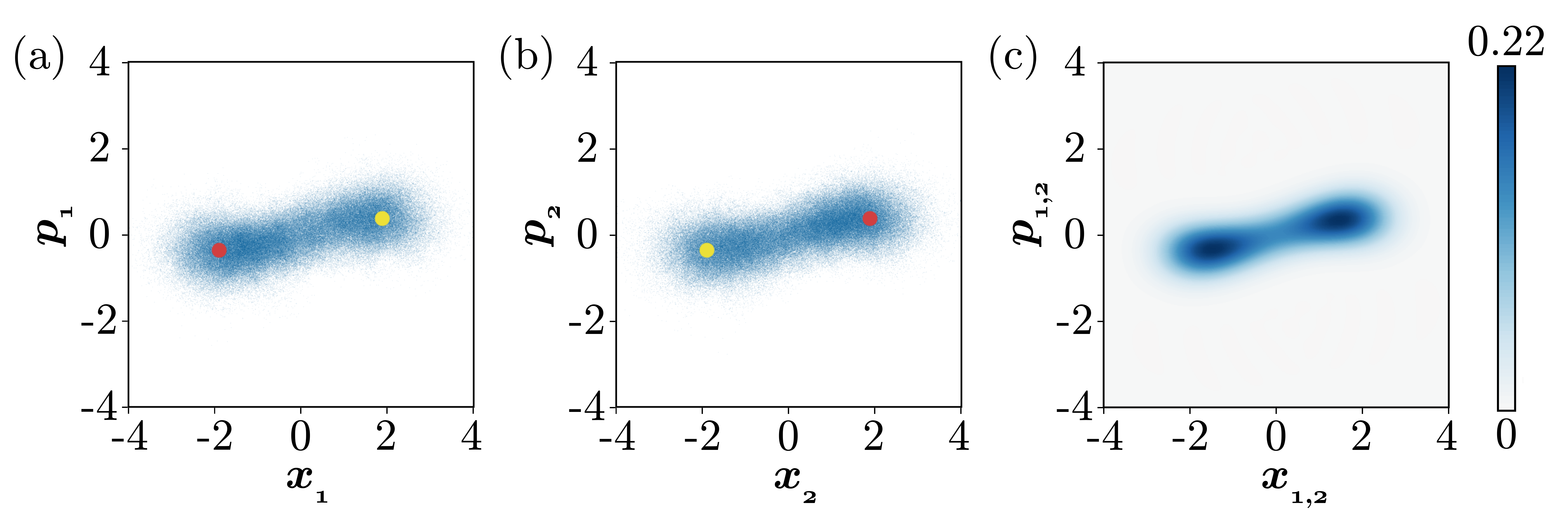}
		\caption{
			\textbf{Scatter plots of stochastic trajectories of two diffusively coupled quantum activator-inhibitor units described by 
				Eq.~(30) in the main text.	
			} (a)~$(x_1,p_1)$ and (b)~$(x_2,p_2)$.
			The semiclassical SDEs of the two coupled units (a, b) have been numerically simulated up to $t = 4000$ with a time interval of $\Delta t = 0.02$ after the initial transient.
			(c) 2D density plot of the stationary Wigner distributions 
			$W(x_1, p_1)$ and $W(x_2, p_2)$ of units $1$ and $2$, which are identical to each other.
			Red and yellow dots in (a,~b) represent stable fixed points 
			of the deterministic classical system.
			The parameters of quantum activator-inhibitor units
			are $\Delta = -0.6, \gamma_{1} = 0.4, \gamma_{2} = 0.1, \theta = \pi$, and $\eta = 0.3$ 
			and the diffusion constants are $D_x = 0.005$ and $D_p = 0.995$ ($D_h = -0.99$ and $D_c = 1$).	
		}
		\label{fig_s1}
	\end{center}
\end{figure}

\clearpage

%

\begin{description}
	\item[Supplementary Movie S1] \mbox{}\\Time evolution of the marginal Wigner distributions of the two units 
	under continuous quantum measurement in the semiclassical regime in Fig.~7.
	\item[Supplementary Movie S2]  \mbox{}\\Time evolution of the marginal Wigner distributions of the two units 
	under continuous quantum measurement in the weak quantum regime in Fig.~8.
	\item[Supplementary Movie S3]  \mbox{}\\Time evolution of the marginal Wigner distributions of the two units 
	under continuous quantum measurement in the strong quantum regime in Fig.~9.
\end{description}
\end{document}